%% file: main.tex
\pdfoutput=1
\documentclass[11pt]{article}

\usepackage{ACL2023}

\usepackage{times}
\usepackage{latexsym}

\usepackage[T1]{fontenc}
\usepackage[utf8]{inputenc}

\usepackage{microtype}

\input{math_commands.tex}

\usepackage[ruled,vlined]{algorithm2e}
\usepackage{hyperref}
\usepackage{comment}
\usepackage{url}
\usepackage{graphicx}
\usepackage{multicol, multirow}
\usepackage{booktabs}
\usepackage{amsthm}
\usepackage{subcaption}
\usepackage{xcolor}
\usepackage{bbm}

\newtheorem{assumption}{Assumption}
\newtheorem{corollary}{Corollary}
\newtheorem{definition}{Definition}
\newtheorem{lemma}{Lemma}
\newtheorem{theorem}{Theorem}

\def\asru/{ASR-U}
\def\wavtovecu/{wav2vec-U}
\def\wavtovectwo/{wav2vec 2.0}
\def\smrm/{symmetric Markov random matrix}
\def\sbp/{small ball probability}
\def\Sbp/{Small ball probability}
\def\whp/{with high probability}
\def\Wlog/{Without loss of generality}
\def\subg/{subgaussian}
\def\debruijn/{de Bruijn}

\title{A Theory of Unsupervised Speech Recognition}

\author{Liming Wang$^1$, Mark Hasegawa-Johnson$^1$ \and Chang D. Yoo$^2$ \\
$^1$University of Illinois Urbana-Champaign \\
$^2$Korea Advanced Institute of Science Technology\\
 \small{\texttt{\{lwang114,jhasegaw\}@illinois.edu}, \texttt{cd\_yoo@kaist.ac.kr}}
}

\begin{document}
\maketitle
\begin{abstract}
\input{abstract.tex}
\end{abstract}

\section{Introduction}
\input{intro}

%\section{Preliminary}
%\input{preliminary}

\section{Problem formulation}
\input{problem_formulation}

\section{Theoretical analysis of \asru/}
\subsection{Learnability of \asru/: a sufficient condition}\label{sec:hmm}
\input{sufficient_condition_asru}

\subsection{Finite-sample learnability of \asru/}\label{sec:finite_sample}
\input{finite_sample_learning}

\subsection{Training dynamic of GAN-based \asru/}
\input{training_dynamic}\label{sec:train_dynamic}

\section{Experiments}
\input{experiments}

\section{Related works}
\input{related_works}

\section{Conclusion}
\input{conclusion}

\section{Limitations}
\input{limitations}

\section*{Acknowledgements}
This work was supported by Institute of Information \& communications Technology Planning \& Evaluation (IITP) grant funded by the Korea government(MSIT) (No.2022-0-00184, Development and Study of AI Technologies to Inexpensively Conform to Evolving Policy on Ethics)

\bibliography{anthology,acl2023}
\bibliographystyle{acl_natbib}

\appendix

\input{appendix}

\input{appendix_reproducibility}
\end{document}

%% file: math_commands.tex
\usepackage{amsmath,amsfonts,bm}

\def\Figref#1{Figure~\ref{#1}}

\def\Secref#1{Section~\ref{#1}}
\def\eqref#1{Eq.~\ref{#1}}
\def\Eqref#1{Eq.~(\ref{#1})}
\def\1{\bm{1}}

\DeclareMathAlphabet{\mathsfit}{\encodingdefault}{\sfdefault}{m}{sl}
\SetMathAlphabet{\mathsfit}{bold}{\encodingdefault}{\sfdefault}{bx}{n}

\def\gE{{\mathcal{E}}}
\def\gF{{\mathcal{F}}}
\def\gG{{\mathcal{G}}}
\def\gH{{\mathcal{H}}}

\def\gN{{\mathcal{N}}}

\def\gW{{\mathcal{W}}}

\def\sN{{\mathbb{N}}}

\def\sR{{\mathbb{R}}}

\def\sX{{\mathbb{X}}}
\def\sY{{\mathbb{Y}}}
\def\sZ{{\mathbb{Z}}}

\newcommand{\Cs}{\mathcal{C}}
\newcommand{\E}{\mathbb{E}}
\newcommand{\Ls}{\mathcal{L}}

\newcommand{\Var}{\mathrm{Var}}

\newcommand{\diag}{\mathrm{diag}}
\DeclareMathOperator{\Tr}{Tr}

%% file: abstract.tex
Unsupervised speech recognition (\asru/) is the problem of learning
automatic speech recognition (ASR) systems from \emph{unpaired}
speech-only and text-only corpora. While various algorithms exist to solve this problem, a theoretical framework is missing to study their properties and addressing such issues as sensitivity to hyperparameters and training instability. In this paper, we proposed a general theoretical framework to study the properties of \asru/ systems based on random matrix theory and the theory of neural tangent kernels. Such a framework allows us to prove various learnability conditions and sample complexity bounds of \asru/. Extensive \asru/ experiments on synthetic languages with three classes of transition graphs provide strong empirical evidence for our theory (code available at \href{https://github.com/cactuswiththoughts/UnsupASRTheory.git}{cactuswiththoughts/UnsupASRTheory.git}).
%\href{https://anonymous.4open.science/r/UnsupASRTheory-4A9E}{here}).

%% file: intro.tex
Unsupervised speech recognition (\asru/) addresses the problem of learning
automatic speech recognition (ASR) systems using \emph{unpaired}
speech-only and text-only corpora. Such a system not only 
significantly reduces the dependency on annotation resources required for
training state-of-the-art ASR system, but it also serves as a bridge between
spoken and written language understanding tasks in the low-resource
settings. Since its first proposal~\cite{Liu2018-asru}, it has seen remarkable
progress, and the current best
system~\cite{Baevski2021-wav2vec-u} has achieved comparable
performance to systems trained with paired data on various
languages. However, there are several ambiguities 
surrounding \asru/, which could impede future evolution of these systems. In particular, prior experiments have found that
training the current state-of-the-art \asru/
model, \wavtovecu/~\cite{Baevski2021-wav2vec-u}, requires careful
tuning over the weights of various regularization losses to avoid
converging to bad local optima and that even despite extensive
regularization weight tuning, \wavtovecu/ may still fail to
converge \cite{Ni-etal-2022-unsup-tts}. Therefore, it is ambiguous whether or when unpaired speech and text data indeed provide
sufficient information for learning an ASR system. Another ambiguity is
whether the success of existing \asru/ models based on generative
adversarial net (GAN)~\cite{Goodfellow2014} is sufficiently explained
by the GAN objective function or requires other
factors, such as randomness in training, quirks in the data used and
careful domain-specific hyper-parameter settings.

This paper provides a theoretical analysis of \asru/ to
clarify the ambiguity surrounding \asru/. First, we prove
learnability conditions and sample complexity bounds that depend on the eigenvalue spacings of the transition probability matrix
of the spoken language. Random matrix theory shows that such
learnability conditions are achievable \whp/. Next, we study the gradient
flow of GAN-based \asru/ and provide conditions under which the
generator minimizing the GAN objective converges to the true
generator. Finally, to verify our theory empirically, we perform GAN-based \asru/ experiments on three classes of synthetic languages. Not only do we observe phase transition phenomena predicted by our theory,  but we achieve stable training with a lower test word error rate by several modifications of the existing state-of-the-art \asru/ system inspired by our theory.

%% file: problem_formulation.tex
\begin{figure}[t]
    \centering
    \includegraphics[width=0.49\textwidth]{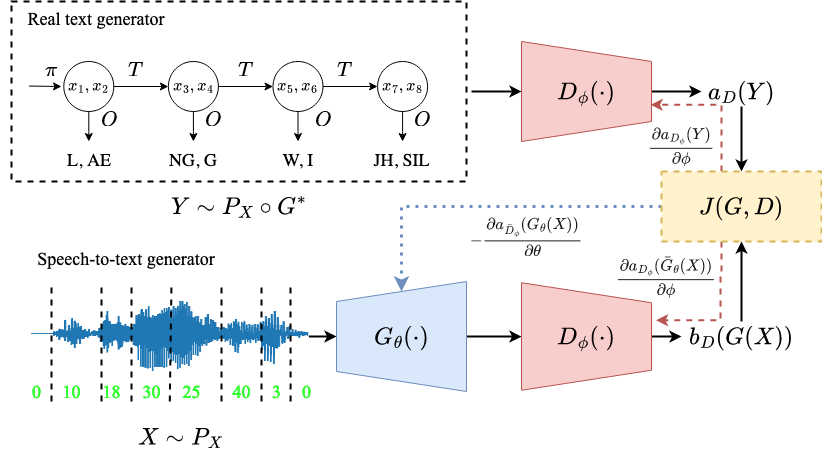}
    \caption{Overview of the ASR-U system for our analysis}
    \label{fig:asr_u}
\end{figure}

% The problem
\paragraph{General formulation}

%Given a random sequence of speech feature vectors or waveform segments $X=[X_1,\cdots,X_L]$ and a random sequence of phoneme labels $Y=[Y_1,\cdots,Y_L]$, which are \emph{unpaired} but are known to have a \emph{matched} distribution in the sense that $P_X$, the pdf of $X$, can be used to generate a distribution $P_Y$, the pmf of $Y$, with some generator $G:\sX\mapsto \sY$ such that $P_X \circ G=P_Y$. Equivalently, there exists a function $G$ such that:

The training data consists of quantized speech sequences
and phoneme label sequences.  The two sequences are
unpaired: no label sequence matches any one of the
speech sequences.  The data are, however, matched in distribution.
Let $P_{X_i}(x)$ and $P_{Y_j}(y)$ be the probability mass functions
(pmfs) of the $i^{\textrm{th}}$ speech vector in a sequence,
$x\in\sX$, and the $j^{\textrm{th}}$ phoneme in a sequence,
$y\in\sY$, respectively: the requirement that they are {\em
  matched in distribution} is the requirement that there exists some
generator function $O:(\sX,\sY)\rightarrow\{0,1\}$ such that
%\begin{align}\label{eq:unsup_asr}
%    \int_{x:G(x)=y} P_X(\mathrm{d}x) = P_Y(y),\forall y\in\sY.
%\end{align}
\begin{align}\label{eq:unsup_asr}
  \sum_{x\in\sX}P_{X_i}(x)O(x,y) = P_{Y_i}(y)
\end{align}
The problem of \asru/ is to find the generator function $O$.

%By the law of total probability, it is easy to see that this also implies that the \emph{$N$-gram} distributions for all $N < L$ of  the generator distribution matches that of the true distributions. The problem of \asru/ is then to find such a generator $G$.

\paragraph{GAN-based \asru/}

%Since the speech set $\sX$ is a set of continuous vectors with unknown and potentially complicated pdf, in reality we cannot observe every possible value of $X$, but instead must estimate $G$ from samples. To address the issue, a GAN can be used to exactly reproduce the empirical distribution of the training dataset with minimum violations of the generator's inductive bias, e.g., with minimum violations of the local spectrotemporal smoothness of the speech features, as encouraged by the generator's CNN front end, as shown in \Figref{fig:asr_u}. It achieves this goal by playing a zero-sum game with another neural network called the discriminator $D$ with the following general utility function:

Eq.~(\ref{eq:unsup_asr}) leverages sequence information to remove
ambiguity: $O$ must be an optimal generator not only for the
position-independent distributions of $X$ and $Y$, but also for their
position-dependent distributions $P_{X_i},P_{Y_i}\forall i\in\sN^0$.  In
reality we cannot observe every possible sequence of speech vectors,
or every possible sequence of phonemes, but instead must estimate $O$
from samples.  To address this issue, a GAN can be used to reproduce
the empirical distribution of the training dataset with minimum error,
subject to the generator's inductive bias, e.g., subject to the
constraint that the function $O$ is a matrix of the form
$O\in\{0,1\}^{\vert\sX\vert\times\vert\sY\vert}$, where
$\vert\sX\vert$ and $\vert\sY\vert$ are the sizes of the alphabets
$\sX$ and $\sY$, respectively.  As shown in \Figref{fig:asr_u}, a GAN
achieves this goal by computing $O$ as the output of a neural network,
$O=G(x,y;\theta)$, and by requiring $G$ to play a zero-sum game with
another neural network called the discriminator $D$ with the following
general utility function:
\begin{multline}\label{eq:gan_objective}
        \min_G \max_D J(G, D) := 
       \E_{Y\sim P_Y}[a(D(Y))] -\\ \E_{X\sim P_X}[b(D(G(X)))].
\end{multline}
For the original GAN~\cite{Goodfellow2014}, $a(D)=\log(\sigma(D))$ and $b(D)=-\log(1-\sigma(D))$, where $\sigma$ is the sigmoid function.
For the Wasserstein GAN~\cite{Arjovsky17}, $D(Y)$ is a Lipschitz-continuous scalar function, and
$a(D)=b(D)=D$.  A maximum mean discrepancy (MMD) GAN~\cite{li2017mmd} minimizes
the squred norm of \Eqref{eq:gan_objective}, where
$D(Y)$ is an embedding into a reproducing kernel Hilbert space (RKHS).  In this paper
we take the RKHS embedding to be the probability mass function of a scalar random variable
$D(Y)$, and assume that the discriminator is trained well enough to
maintain \Eqref{eq:gan_objective}.  
% simplified considerably if the RKHS embedding is the probability mass function of a
% scalar random variable $D(Y)$ non-negative; 
In this situation, the MMD GAN minimizes
Eq.~(\ref{eq:gan_objective}) with $a(D)=b(D)=Y$.
%; for the maximum mean discrepancy (MMD) GAN~\cite{Arjovsky17}, $a(D)=b(D)=D$.
In practice, \Eqref{eq:gan_objective} is optimized by alternatively updating the parameters of the discriminator and the generator using gradient descent/ascent:
\begin{align}
  \phi_{i+1} &= \phi_i + \eta \nabla_{\phi}J(G_{\theta_i}, D_{\phi_i})\label{eq:discrim_gd}\\
  \theta_{i+1}
  &=\theta_i - \nu \nabla_{\theta}J(G_{\theta_i}, D_{\phi_{i+1}}).\label{eq:gen_gd}
\end{align}

% Key questions to answer: When is ASR-U feasible? What makes GAN-based learning unstable for ASR-U
\paragraph{Theoretical questions of \asru/}

%The aforementioned formulation of \asru/ is unfortunately, ill-posed. To see this, note that for any generator $G$ that satisfies \Eqref{eq:unsup_asr}, we can construct a different $G'$ by swapping any subset of speech waveforms that are assigned to $y$ with an equal measure of speech waveforms that are assigned to $y'$ while maintaining \Eqref{eq:unsup_asr}. One important question is then: \emph{what are the necessary and sufficient conditions for \Eqref{eq:unsup_asr} to have a unique solution?} Further, it is well-known that gradient-based training of GAN can be unstable and prior works on \asru/~\cite{Yeh2019-asru,Baevski2021-wav2vec-u} have used various regularization losses to stabilize training. Therefore, another question of practical significance is: \emph{what are the necessary and sufficient conditions for the alternate gradient method as described by in \Eqref{eq:discrim_gd}-(\ref{eq:gen_gd}) to converge to the true generator for \asru/?} In the subsequent sections, we set out to answer these questions.

The aforementioned formulation of \asru/ is ill-posed.  Intuitively,
the function $O$ has finite degrees of freedom
($O\in\{0,1\}^{\vert\sX\vert\times\vert\sY\vert}$), while
\Eqref{eq:unsup_asr} must be valid for an infinite number of
distributions ($P_{X_i}$ and $P_{Y_i}$ for $i\in\sN$), so there is no
guarantee that a solution exists. On the other hand, if the sequence
is unimportant ($P_{X_i}=P_{X_j}\forall i,j\in\sN^0$), then
the solution may not be unique.
One important question is then:
\emph{what are the necessary and sufficient conditions for
  \Eqref{eq:unsup_asr} to have a unique solution?} 
  
  Further, it is well-known that gradient-based training of GAN can be unstable and prior works on \asru/~\cite{Yeh2019-asru,Baevski2021-wav2vec-u} have used various regularization losses to stabilize training. Therefore,
another question of practical significance is: \emph{what are the necessary and sufficient conditions for the alternate gradient method as described by \Eqref{eq:discrim_gd}-(\ref{eq:gen_gd}) to
  converge to the true generator for \asru/?} 
  In the subsequent sections, we set out to answer these questions.

%% file: sufficient_condition_asru.tex
A key assumption of our theory is that the distribution of the speech and the text units can be modeled by a single \emph{hidden Markov model} whose hidden states are $N$-grams of speech units and whose outputs are $N$-grams of text units, as shown in \Figref{fig:asr_u}.

The parameters of this HMM are its initial probability vector, $\pi$,
which specifies the distribution of the first $N$ speech vectors
$X_{0:(N-1)}\in\sX^{N}$, its transition probability matrix $A$, which
specifies the probability of any given sequence of $N$ speech vectors
given the preceding $N$ speech vectors, and its observation
probability matrix, which specifies the distribution of one phone
symbol given one speech vector:
\begin{align*}
\pi &:= P_{X_{0:N-1}}\in\Delta^{\vert\sX\vert^N}\\
A &:= P_{X_{N:2N-1}|X_{0:N-1}}\in\Delta^{\vert\sX\vert^N\times\vert\sX\vert^N}\\
O &:= P_{Y|X} \in \Delta^{|\sX|\times |\sY|},
\end{align*}
where $\Delta^k$ is the $k$-dimensional probability simplex. 

The
first-order Markov assumption is made plausible by the use of N-gram
states, $X_{0:N-1}$, rather than unigram states; with sufficiently
long $N$, natural language may be considered to be approximately
first-order Markov.  The connection between the $N$-gram states and
the unigram observations requires the use of a selector matrix,
$E=\mathbf{1}_{|\sX|^{N-1}}\otimes I_{|\sX|}$, where $\otimes$ denotes
the Kronecker product, thus $P_{X_{kN}}=\pi^{\top}A^kE$, and for multiples
of $N$, \Eqref{eq:unsup_asr} can be written $P_{Y_{kN}}=\pi^{\top}A^kEO$.
It turns out that a crucial feature for a spoken language to be
learnable in an unsupervised fashion is that it needs to be
``complex'' enough such that a simple, symmetric and repetitive graph
is not sufficient to generate the language. This is captured by the
following assumptions on the parameters $A$ and $\pi$.

% we assume additional properties on the submatrices of the acoustic model transition matrix $T$.
%\begin{assumption}\label{assu:eigenvals_of_T}
%There exists an invertible matrix $U\in \sR^{|\sX|^{N-1}\times |\sX|^{N-1}}=[U_1|U_2|\cdots|U_K]$, where the columns of each matrix $U_j$ are eigenvectors with the same eigenvalue and a diagonal matrix $\Lambda=\mathrm{blkdiag}(\Lambda_1,\cdots,\Lambda_K)$, where each matrix $\Lambda_k$ is a diagonal matrix with all diagonal elements equal to the same scalar $\lambda_k$, such that $T=U\Lambda U^{-1}$ with $|\sX|^N \geq K \geq |\sX|$ nonzero eigenvalues $\lambda_{1} > \lambda_{2} >\cdots >\lambda_{K}$, at least $|\sX|$ of which are nonzero.
% Every submatrix of the acoustic model transition probability matrix $T$ in the form $T^{i(1, x_{2:N-1})}:=(T_{i(x,x_{2:N-1}), x'})_{x,x'\in \sX^2}$ is diagonalizable, invertible and the set of eigenvalues formed by such matrices have at least $|\sX|$ members, where $i(x_1,\cdots,x_{N-1})=\sum_{i=1}^{N-1} x_i |\sX|^{i-1}$ is an indexing function that maps $(N-1)$-grams to unique integer indices. 
%\end{assumption}
    
%Less intuitively, we need the following condition for the initial probability.
%\begin{assumption}\label{assu:init_prob}
%The initial probability satisfies $\pi^\top u_{jk} \neq 0, U_j = (u_{jk})_{1\leq k\leq K_j}, 1\le j\le |X|$, $1\le k\le N_j$, where $U_j=[u_{j1}|\cdots|u_{jN_j}]$.
%\end{assumption}
% we assume additional properties on the submatrices of the acoustic model transition matrix $T$.

\begin{assumption}\label{assu:eigenvals_of_T}
There exists an invertible matrix $U\in \sR^{|\sX|^{N-1}\times
|\sX|^{N-1}}=[U_1|U_2|\cdots|U_K]$, where the columns of each matrix
$U_j=[u_{j1}|\cdots|u_{jN_j}]$ are eigenvectors with the same
eigenvalue and a diagonal matrix
$\Lambda=\mathrm{blkdiag}(\Lambda_1,\cdots,\Lambda_K)$, where each
matrix $\Lambda_k$ is a diagonal matrix with all diagonal elements
equal to the same scalar $\lambda_k$, such that $A=U\Lambda U^{-1}$
with $|\sX|^N \geq K \geq |\sX|$ nonzero eigenvalues $\lambda_{1}
> \lambda_{2} >\cdots >\lambda_{K}$.
\end{assumption}
    
\begin{assumption}\label{assu:init_prob}
For at least $\vert\sX\vert$ values of $j$, there is at least one $k$
s.t.~$\pi^\top u_{jk}\neq 0$.
\end{assumption}
    
With Assumptions~\ref{assu:eigenvals_of_T} and~\ref{assu:init_prob}, we can consider the following algorithm: First, we construct the following matrices
\begin{align}
    P^X := \begin{bmatrix}
    P_{X_{0}}^\top\\
    P_{X_{N}}^\top\\
    \vdots\\
    P_{X_{(L-1)N}}^\top
    \end{bmatrix}&, P^Y := \begin{bmatrix}
    P_{Y_{0}}^\top\\
    P_{Y_{N}}^\top\\
    \vdots\\
    P_{Y_{(L-1)N}}^\top
    \end{bmatrix},\label{eq:asymp_px_py_def}
\end{align}
Then, $O$ satisfies the following matrix equation
\begin{align}\label{eq:asymp_asr_u}
    P^{X}O &= P^Y.
\end{align}
%Note that \eqref{eq:asymp_asr_u} has a unique solution if and only if $P^X$ has full column rank. Note that, because of the relationship between X and Y assumed in \eqref{eq:unsup_asr}, \eqref{eq:asymp_asr_u} is true even when $P^X$ has more rows than columns ($L > |X|$).  The generator function $G(x)$ in \eqref{eq:asymp_asr_u} (equivalently: the binary matrix $O$ in \eqref{eq:asymp_asr_u}) is unique if and only if $P^X$ has full column rank. The following theorem proves that this is indeed the case under our assumptions.
The binary matrix $O$ in \Eqref{eq:asymp_asr_u} is unique if and only if $P^X$ has full column rank. The following theorem proves that this is indeed the case under our assumptions.

\begin{theorem}\label{thm:asymp_unsup_asr}
Under Assumptions~\ref{assu:eigenvals_of_T} and~\ref{assu:init_prob}, $P^X$ has full column rank and perfect \asru/ is possible. Further, the true phoneme assignment function is $O = P^{X+}P^Y$, where $P^{X+}=(P^{X\top}P^X)^{-1}P^{X\top}$ is the left-inverse of $P^{X}$.
\end{theorem}
    
Further, if we measure how far the matrix $P^X$ is from being singular by its \emph{smallest} singular value defined as
\begin{align*}
    \sigma_{\min}(P^X) := \min_{v \in \sR^{|\sX|}}\frac{\|P^X v\|_2}{\|v\|_2},
\end{align*}
we can see that $P^X$ becomes further and further away from being singular as the sequence length $L$ gets larger. An equivalent result for a different purpose has appeared in the Theorem 1 of \cite{Bazan2000}.

%\begin{proposition}\label{prop:least_singular_val_unsup_asr}
%Under Assumption~\ref{assu:eigenvals_of_T} and for simplicity assuming the number of distinct eigenvalues $K=|\sX|$  for $T$, then we have 
%     \begin{multline}
%         \sigma_{\min}(P^X) \geq \\ \frac{\left(|\det(V)||\sX|!\right)^{1/|\sX|}\delta_{\min}^{\frac{|\sX|-1}{2}}}{\kappa(P^X_{1:|\sX|})}\sqrt{\sum_{l=0}^{L-1}\lambda_{\min}^{2|\sX|l}(T)},
%    \end{multline}
%where $\delta_{\min}:= \min_{i\neq j}|\lambda_i(T)-\lambda_j(T)|$, $\lambda_{\min}(A)$ is the smallest eigenvalue of square matrix $A$,  $P^X_{1:|\sX|}$ is the first $|\sX|$ row of $P^X$ and $\kappa(A)$ is the condition number of the matrix $A$.
%    \end{proposition}

\begin{lemma}\label{prop:least_singular_val_unsup_asr}
Under Assumptions~\ref{assu:eigenvals_of_T} and~\ref{assu:init_prob} and
for simplicity assuming the number of distinct eigenvalues $K=|\sX|$
for $T$, then we
have
\begin{multline}
\sigma_{\min}(P^X) \geq \\
%\frac{\left(|\det(V)||\sX|!\right)^{1/|\sX|}\delta_{\min}^{\frac{|\sX|-1}{2}}}{\kappa(P^X_{1:|\sX|})}
%\sqrt{\sum_{l=0}^{L-1}\lambda_{\min}^{2|\sX|l}(T)},
\frac{\delta_{\min}^{(|\sX|-1)/2|\sX|}\sum_{l=0}^{L-|\sX|-1}\lambda_{\min}^{2l}(A)}
{\kappa(V_{|\sX|}(\lambda_{1:|\sX|}))}
\min_j\|\hat{r}_j\|
\end{multline}
where $\delta_{\min}:= \min_{i\neq j}|\lambda_i(A)-\lambda_j(A)|$,
$\lambda_{\min}(A)$ is the smallest eigenvalue of square matrix $A$,
$\kappa(V_{|\sX|}(\lambda_{1:|\sX|}))$ is the condition number of the
square Vandermonde matrix created from eigenvalues
$\lambda_1(A),\ldots,\lambda_{|\sX|}(A)$, $r_j=\pi^TU_j\Omega_j^\top E$,
and $\Omega_j^\top$ is the set of rows of $U^{-1}$ corresponding to
eigenvalue $\lambda_j(A)$, after orthogonalizing them from every other
block of rows, i.e., $U^{-1}=L[\Omega_1|\cdots|\Omega_K]^T$ such that $L$ is
lower-triangular, and the blocks $\Omega_i$ and $\Omega_j$ are orthogonal.
\end{lemma}

Next, we will show that Assumption~\ref{assu:eigenvals_of_T} can be easily met using random matrix arguments.

%Instead of analyzing the behaviour of overlapping $N$-grams, we find it much easier to analyze \emph{non-overlapping} $N$-grams, since this can avoid various complications due to the underlying Markov transition dynamic being non-reversible and allow us to apply classical results from random matrix theory more conveniently. As a result, it remains an open question to perform analysis along the same line for a non-reversible Markov chain, such as one containing many cycles.

%% file: finite_sample_learning.tex
    \paragraph{Matched setup} Now we show that the requirement for distinct eigenvalues is a mild one as it can easily be satisfied with \emph{random} transition matrices. According to such a result, ASR-U is feasible with high probability in the (empirically) \emph{matched} setting commonly used in the ASR-U literature, where the \emph{empirical} generated and true distributions can be matched exactly by some generator in the function class~\cite{Liu2018-asru}. Our proof relies crucially on the seminal work of \cite{NguyenTaoVu2015} on eigenvalue gaps of symmetric random matrices with independent entries.
%\begin{theorem}{(Theorem 2.6 of \cite{NguyenTaoVu2015})}\label{thm:eigenval_gap_symmetric_matrix}
%    For any real symmetric matrix of the form $W_n = F_n + X_n$, where $F_n \in \sR^{n\times n}$ is a deterministic symmetric matrix with eigenvalues of order $O(n^\gamma)$ and $X_n$ is a zero-mean, subgaussian random matrix, we have for any $C > 0$, there exists $B > \max\{5\gamma' C + 3\gamma' + 3, 3\gamma'+3\}$ s.t.
 %   \begin{align*}
 %       \max_{1\leq i \leq n-1}\Pr[|\lambda_i-\lambda_{i+1}| \leq n^{-B}]\leq n^{-C},
 %   \end{align*}
 %   with probability at least $1-O(\exp(-\alpha_0 n))$ for some $\alpha_0 > 0$ dependent on $B$ and $\gamma'=\max\{\gamma, 1/2\}$.
%\end{theorem}
    
    In the context of ASR-U, it is of particular interest to study the eigenvalue gaps of a Markov random matrix, which unlike the symmetric case, is asymmetric with correlated entries. Fortunately, by modifying the proof for Theorem 2.6 of \cite{NguyenTaoVu2015}, we can show that if the language model belongs to a special but rather broad class of Markov random matrices defined below and the states are \emph{non-overlapping} $N$-gram instead of the more common overlapping ones, it should have at least $|\sX|$ distinct eigenvalues with minimum spacing depending on $|\sX|$ and the $N$ for the $N$-gram.

    \begin{definition}{(\smrm/)}
    A \smrm/ is a matrix of the form $A := D^{-1}W$, where the \emph{adjacency matrix} $W$ is a real, symmetric random matrix with positive entries and bounded variance and $D$ is a diagonal matrix with $d_{ii}=\sum_{j} W_{ij} > 0$.
    \end{definition}
    Intuitively, a \smrm/ is the transition matrix for a \textit{reversible} Markov chain formed by normalizing edge weights of a weighted, undirected graph.
    \begin{theorem}{(simple spectrum of \smrm/)}\label{thm:simple_spectrum}
    Let $A_n = D_n^{-1}W_n \in \sR^{n\times n}$ be a real \smrm/ with adjacency matrix $W_n$. Further, suppose $W_n=F_n+X_n$, where $F_n$ is a deterministic symmetric matrix with eigenvalues of order $n^{\gamma}$ and $X_n$ is a
    symmetric random matrix of zero-mean, unit variance sub-Gaussian random variables. Then
    %zero-mean, symmetric \subg/ random matrix,
    we have for any $C > 0$, there exists $B > 4\gamma' C+7\gamma'+1$ such that
    \begin{align*}
        \max_{1\leq i \leq n-1}\Pr[|\lambda_i-\lambda_{i+1}| \leq n^{-B}]\leq n^{-C},
    \end{align*}
    with probability at least $1-O(\exp(-\alpha_0 n))$ for some $\alpha_0 > 0$ dependent on $B$ and $\gamma'=\max\{\gamma, 1/2\}$.
    \end{theorem}
    %MH: following paragraph is not necessarily true, because $d_{ii}$ may be larger than O(n)
    %Note that compared to Theorem~\ref{thm:eigenval_gap_symmetric_matrix}, the eigenvalue gap for a \smrm/ is in average smaller by at least an order of $n^{-1}$, which is because the normalization factor $d_{ii}$ is of order $O(n)$.
    \begin{corollary}\label{col:simple_spectrum}
    Suppose the speech feature transition probability is a \smrm/ $A := D^{-1}W$ with entries $W_{ij}\sim \mathrm{Uniform}(0, 2\sqrt{3})$ and $D$ is a diagonal matrix with $d_{ii}=\sum_{j}W_{ij}$. Then for any $\epsilon > 0$, there exists $\alpha_0 > 0$ such that with probability at least $1-O\left(|\sX|^{-CN}+\exp\left(-\alpha_0|\sX|^N\right)\right)$, the transition probability matrix $A$ has $|\sX|^N$ distinct eigenvalues with minimum gap $|\sX|^{-BN} > 0$.
    \end{corollary}
    The proof of Theorem~\ref{thm:simple_spectrum} and Corollary~\ref{col:simple_spectrum} are presented in detail in the Appendix~\ref{appendix:proofs_finite_sample}. 
    
    \paragraph{Unmatched setup} In the finite-sample, unmatched setup, the empirical distribution of the fake text data generated by the GAN does not necessarily match the empirical distribution of the true text data. Assuming the discriminator is perfect in the sense that it
    maintains \Eqref{eq:gan_objective} non-negative, and assuming $D(Y)$ is a scalar random variable, then minimizing \Eqref{eq:gan_objective} is equivalent to minimizing a
    %maximize \Eqref{eq:gan_objective} with respect to any generator $G$, then the GAN boils down to finding an assignment matrix $O$ such that certain
    \emph{divergence measure} $d(\cdot, \cdot)$, between the empirical text distribution, $P^Y$, and the text distribution generated by $O_x(y)=\hat{P}_{Y|X}(y|x)$:
    \begin{align}\label{eq:discrete_perfect_discriminator}
        %\min_{O\in \Delta^{|\sX|\times |\sY|}} d^{\gamma}(\hat{P}_Y,\hat{P}_{X} \circ O),
        \min_{O\in \Delta^{|\sX|\times |\sY|}} d^{\gamma}(P^Y,P^{X} O),
    \end{align}
    where $\gamma > 0$. For example, for the original GAN, $d(\cdot,\cdot)$ is the Jensen-Shannon distance and for the MMD GAN, $d(\cdot, \cdot)$ is the $L_\gamma$ distance between the expectations $\mathbb{E}[D(Y)]$ under the distributions $P^Y$ and $P^XO$.  
    % In both cases, Theorem 1 requires \Eqref{eq:discrete_perfect_discriminator} to be the distance between distributions over random sequences, $y=[y_0,\ldots,y_{N(L-1)}]$ of length at least $L$. 
    In both cases, however, \Eqref{eq:discrete_perfect_discriminator} can be minimized using a \emph{decomposable} discriminator 
    % under the condition that the decomposable discriminator is in the domain space of $(P^Y-P^XO)$, where the decomposable discriminator is 
    defined to be:
    \begin{align}\label{eq:decomposable_discrim}
       \E_{P_Y}[a(D(Y))] &= \sum_{l=0}^{L-1} \E_{P_{Y_l}}[a(D_l(Y_l))]\\
       \E_{P_X}[b(D(G(X)))] &= \sum_{l=0}^{L-1}\E_{P_{X_l}}[b(D_l(G_l(X))],
    \end{align}
    %L2 distance, with $\gamma=2$ in both cases. In the case when $d$ is a distance metric, by Jensen's inequality, we then have 
    %\begin{multline}\label{eq:distance_bound}
    %    d^{\gamma}(\hat{P}_Y,\hat{P}_{X} \circ O) \geq\\ L^\gamma|\sX|^{\gamma(L-1)}d^\gamma\left(\hat{P}^Y,\hat{P}^X  O\right),
    %\end{multline}
    %where equality can be achieved when the discriminator is \emph{decomposable} in the sense that
    %\begin{align}\label{eq:decomposable_discrim}
    %    a_D(y) &= \sum_{l=1}^L a_{D_l}(y_l)\\
    %    b_D(y) &= \sum_{l=1}^L b_{D_l}(y_l),
    %\end{align}
    with \emph{components} $D_l:|\sY|\mapsto \sR,l=1,\cdots,L$. Under the assumption that $D$ is decomposable and that the MMD GAN is used, we have the following sample complexity bound on perfect \asru/.
    % Since the text data $Y_{1:N}$ is discrete in nature, we can estimate $P_{Y_{1:N}}$ to be arbitrarily close to the true distribution \emph{uniformly} given enough samples. As a result, we have
    % \begin{align}
    %   \|\hat{P}^X-P^X\|_1 &\leq \epsilon_X,\\
    %   \|\hat{P}^Y-P^Y\|_1 &\leq \epsilon_Y,
    % \end{align}
    % for the number of speech $N$-grams $n_X$ and the number of text $N$-grams $n_Y$ large enough. Suppose the model is able to minimize \eqref{eq:discrete_perfect_discriminator}, the empirical risk minimizer (ERM) will need to satisfy
    % \begin{align}\label{eq:finite_sample_asr_u}
    %    \hat{P}^X \hat{O} = \hat{P}^Y. 
    % \end{align}
    \begin{theorem}\label{thm:discrete_gan_unsup_asr}
    The empirical risk minimizer (ERM) of \Eqref{eq:discrete_perfect_discriminator} recovers the true assignment $O$ perfectly from $n^X$ speech frames and $n^Y$ text characters with probability $1-2\delta$ if 
    \begin{align*}
        \sigma_{\min}(P^X) \geq
        &\sqrt{\frac{4L|\mathbb{Y}|(n^X+n^Y)+L|\mathbb{X}|n^X}{n^Xn^Y}}+\\
        &10\sqrt{\frac{L\log \frac{1}{\delta}}{n^X\wedge n^Y}},
        %\sigma_{\min}(P^X) \geq \frac{3}{\sqrt{n^X}}+\frac{2}{\sqrt{n^Y}}+10\sqrt{\frac{L\log \frac{1}{\delta}}{n^X\wedge n^Y}},
    \end{align*}
    where $n^X\wedge n^Y := \min\{n^X, n^Y\}$.
    \end{theorem}

%% file: training_dynamic.tex
\begin{figure*}[ht]
    \centering
    \begin{subfigure}{0.26\textwidth}
    \includegraphics[width=0.9\textwidth]{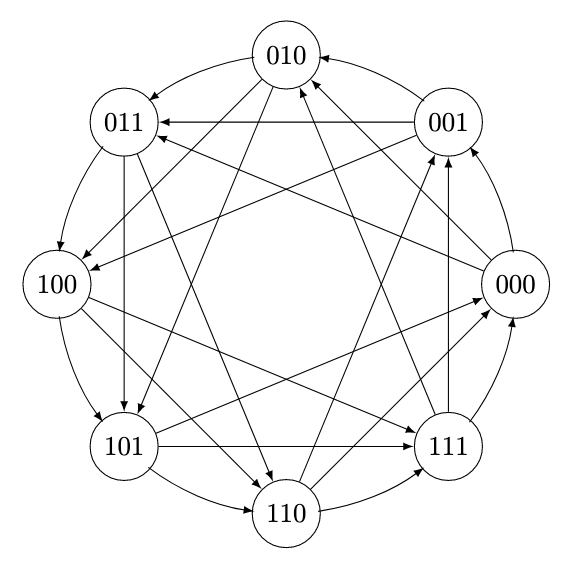}
    \caption{Circulant graph}
    \label{fig:circulant_graph}
    \end{subfigure}
    \begin{subfigure}{0.45\textwidth}
    \includegraphics[width=0.9\textwidth]{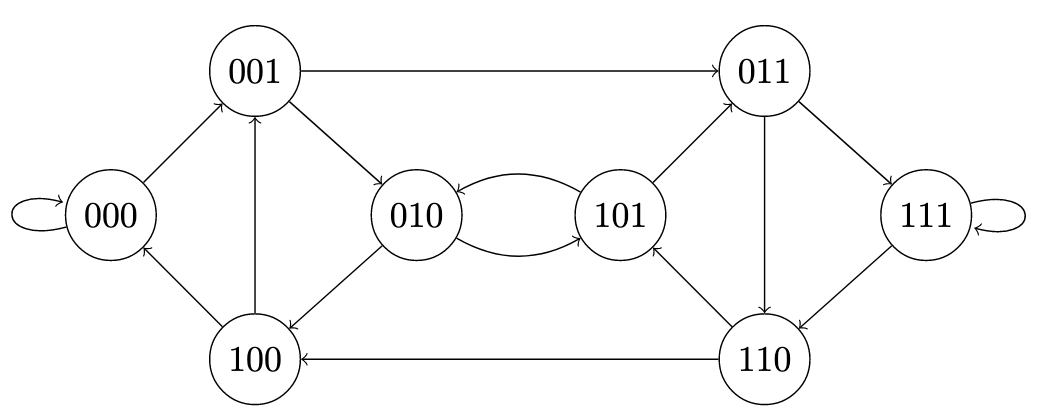}
    \caption{De Bruijn graph}
    \label{fig:debruijn_graph}
    \end{subfigure}
    \begin{subfigure}{0.26\textwidth}
    \includegraphics[width=0.9\textwidth]{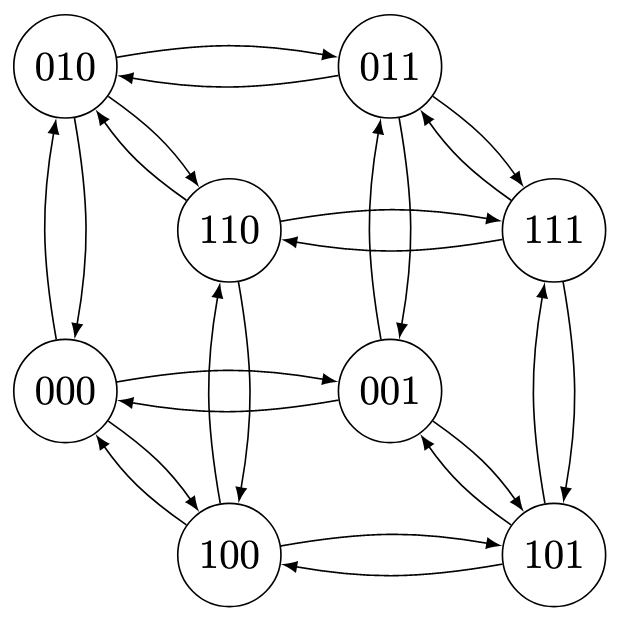}
    \caption{Hypercube}
    \label{fig:hypercube_graph}
    \end{subfigure}
    \caption{Various types of Markov transition graphs}
    \label{fig:graphs}
\end{figure*}
% \textcolor{red}{Discuss about the finite-sample case, the fact that the proof does not rely on x to be discrete}
So far, we have assumed the GAN training is able to find the optimal parameters for the discriminator and the generator. However, there is no guarantee that this is indeed the case with gradient updates such as \Eqref{eq:discrim_gd}. To analyze the behaviour of the GAN training dynamic for \asru/, we follow prior works on neural tangent kernel (NTK)~\cite{Jacot2018-ntk} to focus on the \emph{infinite-width, continuous-time} regime, or NTK regime, where the generator and the discriminator are assumed to be neural networks with an infinite number of hidden neurons trained with gradient descent at an infinitely small learning rate. Though highly idealized, studying such a regime is practically useful as results from this regime can often be converted to finite-width, discrete-time settings (See, e.g., \cite{Du2019}). 

For simplicity, denote $f_\tau:=D_{\phi_\tau}$ and $g_t:=G_{\theta_t}$ and define $\Ls_t(f):=J(g_t, f)$, then in the NTK regime, between each generator step, the training dynamic of the discriminator can be described by the following partial differential equation (PDE):
\begin{align}\label{eq:pde_discrim_gd}
\partial_\tau \phi_\tau = \nabla_{\phi_\tau}\Ls_t(f_\tau).
\end{align}
Let $f_{P_t}^*=\lim_{\tau\rightarrow\infty} f_\tau$ be the limit of \Eqref{eq:pde_discrim_gd}. If the limit exists and is unique, the generator loss is well-defined as $\Cs_{t}(g_t) := J(g_t, f_{P_t}^*)$. Note that the output of the \asru/ generator is discrete, which is not a differentiable function per se, but we can instead directly parameterize the \emph{generated text distribution} as $P_{g_t} := P_X \circ O_t$ for some softmax posterior distribution $O_t$:
\begin{align}\label{eq:gen_param}
    O_{t, x}(y) := \prod_{l=1}^L \frac{\exp(h_{\theta, y_l}(x_l))}{\sum_{y_l'}\exp(h_{\theta, y_l'}(x_l))},
\end{align}
where $h_{\theta}$ is a neural network, and is assumed to be one layer in our analysis, though it can be extended to multiple layers with slight modifications using techniques similar to those in~\cite{Du2019}.

Using such a generator, the generator dynamic can be then described by the following PDE:
\begin{align}\label{eq:pde_gen_gd}
    %\partial_t \theta_t &= -\sum_{y\in \sY^L}b_{f^*_{g_t}}(y)\nabla_{\theta_t}P_{g_t}(y),
    \partial_t \theta_t &= \sum_{y\in \sY^L}b(f^*_{g_t}(y))\nabla_{\theta_t}P_{g_t}(y),
\end{align}
where the right-hand side is the term in the gradient of $\Cs_t$ with respect to $\theta_t$ ignoring the dependency of the discriminator $f_{g_t}^*$. Define the NTKs of the discriminator and the generator (distribution) as 
\begin{align}
    K_{f_\tau}(y, y') &= \E_{\phi_0 \sim \gW}\left[\frac{\partial f_\tau(y)}{\partial \phi_\tau}^\top\frac{\partial f_\tau(y')}{\partial \phi_\tau}\right]\label{eq:ntk_discrim_pd}\\
    K_{g_t}(y, y') &= \E_{\theta_0 \sim \gW} \left[\frac{\partial P_{g_t}(y)}{\partial \theta_t}^\top\frac{\partial P_{g_t}(y')}{\partial \theta_t}\right],
\end{align}
where $\gW$ is the initialization distribution (usually Gaussian). 

Note that the NTKs are $|\sY|^L \times |\sY|^L$ matrices for \asru/ due to the discrete nature of the generator. A key result in~\cite{Jacot2018-ntk} states that as the widths of the hidden layers of the discriminator and generator go to infinity, $K_{f_{\tau}} \rightarrow K_D, K_{g_t}\rightarrow K_G$ stay constant during gradient descent/ascent and we have
\begin{align}
    \partial_\tau f_\tau = K_D& \left(\diag(P_Y)\nabla_{f_\tau}a\right.\nonumber\\
    &\left.-\diag(P_{g_t})\nabla_{f_\tau}b\right), \label{eq:ntk_pde_discrim_gd}\\
    \partial_t P_{g_t} = K_G&\mathbf{b}_{f_{g_t}}\label{eq:ntk_pde_gen_gd},
    %\partial_\tau f_\tau = K_D& (\diag(P_Y)\nabla_{f_\tau} \mathbf{a}_{f_\tau}-\nonumber\\
    %&\diag(P_{g_t})\nabla_{f_\tau} \mathbf{b}_{f_\tau}), \label{eq:ntk_pde_discrim_gd}\\
    %\partial_t P_{g_t} = K_G&\mathbf{b}_{f_{g_t}}\label{eq:ntk_pde_gen_gd},
\end{align}
where $\nabla_f\{a,b\}=\left[\frac{\partial\{a,b\}(f(y))}{\partial f(y)}\right]_{y\in\mathbb{Y}^L}$ and
%where $\mathbf{a}_{f} = (a_f(y))_{y\in \sY^L},\,
$\mathbf{b}_{f} = (b_f(y))_{y\in \sY^L}$. 

However, \Eqref{eq:ntk_pde_discrim_gd}-(\ref{eq:ntk_pde_gen_gd}) is in general highly nonlinear and it remains an open problem as to their convergence properties. Instead, we focus on the case when the discriminator $f_t$ is decomposable with components $f_{t, l}, l=1,\cdots,L$, and simplify \Eqref{eq:ntk_pde_discrim_gd} and \Eqref{eq:ntk_pde_gen_gd} into PDEs involving only samples at a particular time step:
\begin{align}
    \partial_{\tau}f_{\tau, l}=& K_{D, l}\left(\diag(P^Y_l)\nabla_{f_{\tau, l}}\mathbf{a}_{f_{\tau, l}}\right.\nonumber\\
    &-\left.\diag(P^{g_t}_l)\nabla_{f_{\tau, l}}\mathbf{b}_{f_{\tau, l}}\right),\label{eq:ntk_pde_decomp_discrim_gd}\\
    \partial_{t}O_{t, x}^\top =& \sum_{l=1}^L P^X_l(x)K_{O_{t, x}}\mathbf{b}_{f_{g_t, l}},\label{eq:ntk_pde_decomp_gen_gd}
\end{align}
for all $l=1,\cdots,L,\,x\in\sX$ in terms of the \emph{stepwise} NTKs defined as:
\begin{align*}
    K_{D,l}(y, y') &:= \E_{\phi_0\sim \gW}\left[\frac{\partial f_{\tau}(y)}{\partial \phi_\tau}^\top \frac{\partial f_{\tau}(y')}{\partial \phi_\tau}\right]\\
    K_{O_{t,x}}(y, y') &:= \E_{\theta_0\sim \gW}\left[\frac{\partial O_{t, x}(y)}{\partial \theta_\tau}^\top \frac{\partial O_{t, x}(y')}{\partial \theta_\tau}\right].
\end{align*}

We further focus on the special case that $f_{\tau, l}$ is parameterized by a two-layer neural network with ReLU activation, though the framework can be extended to network of arbitrary depths:
\begin{align}\label{eq:discrim_dnn}
    f_{\tau, l}(y) = \lim_{m\rightarrow \infty}\frac{1}{\sqrt{m}}\sum_{r=1}^m v^l_r\max\{W^l_{ry}, 0\}.
\end{align}
In this case, under mild regularity conditions, we can show that the generator trained with the alternate gradient method minimizes \Eqref{eq:discrete_perfect_discriminator}, which under the same condition as in \Secref{sec:finite_sample}, implies \asru/ is feasible.
\begin{theorem}\label{thm:train_dynamic}
Suppose the following assumptions hold:
\begin{enumerate}
    \item The discriminator is decomposable and parameterized by \Eqref{eq:discrim_dnn}, whose parameters are all initialized by standard Gaussian variables;
    \item The generator is linear before the softmax layer;
    \item The GAN objective is MMD;
    \item The linear equation $P^XO=P^Y$ has at least one solution.
\end{enumerate}
 Then we have for any solution $O_t$ of \Eqref{eq:ntk_pde_decomp_gen_gd}, $\lim_{t\rightarrow \infty} P^X O_t = P^Y$.
\end{theorem}

%% file: experiments.tex
\begin{figure*}[ht]
    \centering
    \begin{subfigure}{0.32\textwidth}
    \includegraphics[width=0.99\textwidth]{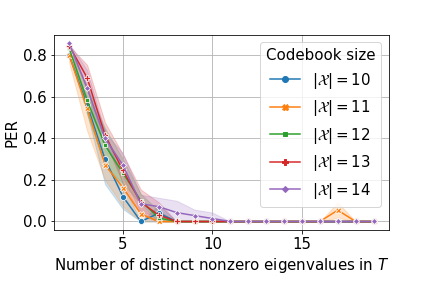}
    \caption{Circulant graph}
    \label{fig:circulant_phase_transition}
    \end{subfigure}
    \begin{subfigure}{0.32\textwidth}
    \includegraphics[width=0.99\textwidth]{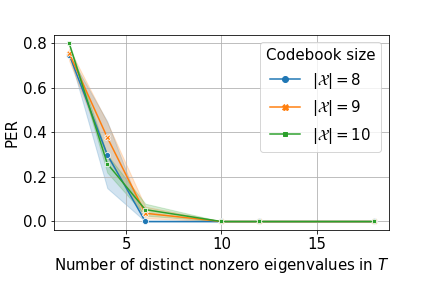}
    \caption{De Bruijn graph}
    \label{fig:debruijn_phase_transition}
    \end{subfigure}
    \begin{subfigure}{0.32\textwidth}
    \includegraphics[width=0.99\textwidth]{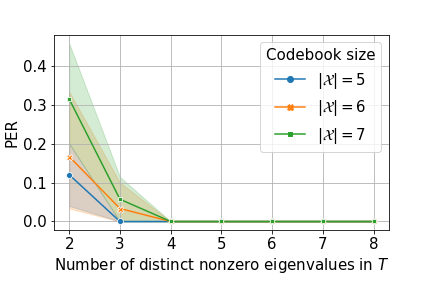}
    \caption{Hypercube}
    \label{fig:hypercube_phase_transition}
    \end{subfigure}
    \caption{Asymptotic \asru/ PER vs number of distinct nonzero eigenvalues for various Markov transition graphs}
    \label{fig:asymp_asru}
\end{figure*}
\begin{figure*}[ht]
    \centering
    \begin{subfigure}{0.32\textwidth}
    \includegraphics[width=0.99\textwidth]{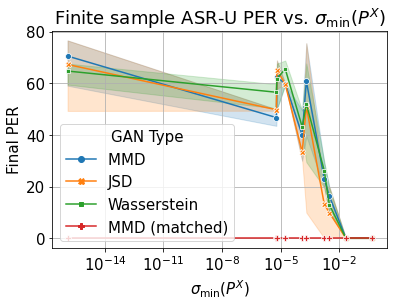}
    \caption{Circulant graph}
    \label{fig:circulant_per_vs_sigma}
    \end{subfigure}
    \begin{subfigure}{0.32\textwidth}
    \includegraphics[width=0.99\textwidth]{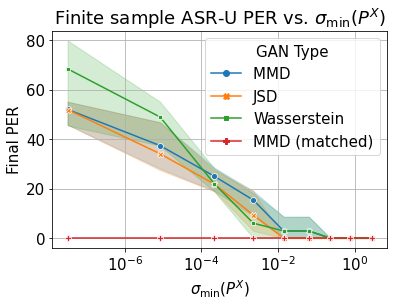}
    \caption{De Bruijn graph}
    \label{fig:debruijn_per_vs_sigma}
    \end{subfigure}
    \begin{subfigure}{0.32\textwidth}
    \includegraphics[width=0.99\textwidth]{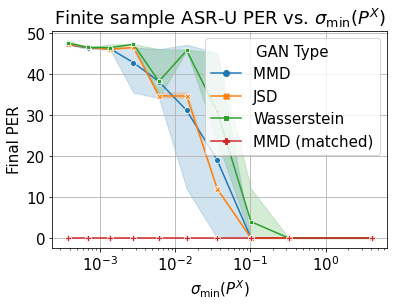}
    \caption{Hypercube}
    \label{fig:hypercube_per_vs_sigma}
    \end{subfigure}
    \caption{Finite-sample \asru/ PER vs $\sigma_{\min}(P^X)$ for various Markov transition graphs}
    \label{fig:finite_sample_asru}
\end{figure*}
\begin{figure}
    \centering
    \includegraphics[width=0.47\textwidth]{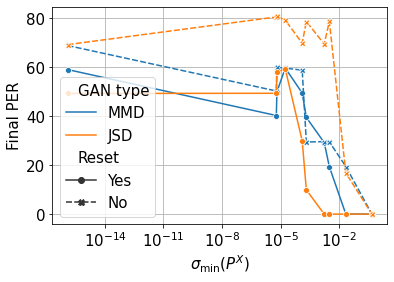}
    \caption{Effect of discriminator resetting at every update}
    \label{fig:discrim_reset}
\end{figure}
\begin{figure}[ht]
    \centering
    \includegraphics[width=0.47\textwidth]{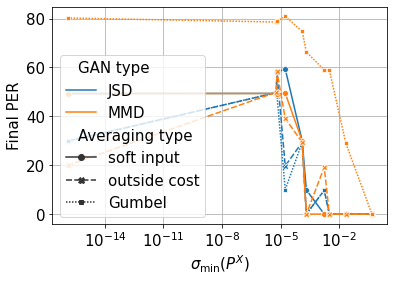}
    \caption{Effect of different type of averaging for the generator}
    \label{fig:gen_avg}
\end{figure}
\paragraph{Synthetic language dataset} To allow easy control of the eigenvalue spacings of the transition matrix $T$ and thus observe the phase transition phenomena predicted by our theory, we design six synthetic languages with HMM language models as follows. First, we create the HMM transition graph by treating non-overlapping \emph{bigrams} as hidden states of the HMM. The hidden state of the HMM will henceforth be referred to as the ``speech unit'', while the observation emitted by the HMM will be referred to as the ``text unit''. For the asymptotic \asru/, we control the number of eigenvalues of the Markov transition graph by varying the number of disjoint, identical subgraphs. The number of distinct eigenvalues of the whole graph will then be equal to the number of eigenvalues of each subgraph. For the finite sample setting, we instead select only Hamiltonian graphs and either gradually decrease the degrees of the original graph to its Hamiltonian cycle or interpolate between the graph adjacency matrix and that of its Hamiltonian cycle. Thus, we can increase $\sigma_{\min}(P^X)$ by increasing $w$. For both the subgraph in the former case and the Hamiltonian graph in the latter, we experiments with circulant, \debruijn/ graphs~\cite{deBruijn1946} and hypercubes, as illustrated in \Figref{fig:graphs}. Next, we randomly permute the hidden state symbols to form the true generator mapping from the speech units to text units. To create matched speech-text data, we simply sample matched speech and text unit sequences using a single HMM. For unmatched datasets, we sample the speech and text data independently with two HMMs with the same parameters. Please refer to Appendix~\ref{appendix:reproducibility} for more details.

% \textcolor{red}{TODO Adding a plot of the eigenvalue distributions and eigengap of smrm}

\paragraph{Model architecture} For finite-sample \asru/, we use \wavtovecu/~\cite{Baevski2021-wav2vec-u} with several modifications. In particular, we experiment with various training objectives other than the Jensen-Shannon (JS) GAN used in the original \wavtovecu/, including the Wasserstein GAN~\cite{Liu2018-asru} and the MMD GAN. All additional regularization losses are \emph{disabled}.
Moreover, we experimentally manipulate two hyperparameters: (1) the averaging strategy used by the generator, and (2) whether to \emph{reset} the discriminator weights to zero at the beginning of each discriminator training loop. More details can be found in Appendix~\ref{appendix:reproducibility}.

\paragraph{Phase transition of PER vs. eigenvalue gaps: asymptotic case} The phoneme error rate (PER) as a function of the number of eigenvalues of $A$ for the asymptotic \asru/ on the synthetic datasets are shown in Figure~\ref{fig:asymp_asru}. For all three graphs, we observe clear phase transitions as the number of eigenvalues exceeds the number of speech units, and an increase of the number of distinct, nonzero eigenvalues required for perfect \asru/ as the number of speech units increases.

\paragraph{Phase transition of PER vs. eigenvalue gaps: finite-sample case} The PER as a function of the least singular value $\sigma_{\min}(P^X)$ for the finite-sample \asru/ on the synthetic datasets are shown in \Figref{fig:finite_sample_asru}. As we can see, the \asru/ exhibit the phase transition phenomena in all three graphs, albeit with differences in the critical point and their rate of approaching the perfect \asru/ regime. While the PER generally decreases as $\sigma_{\min}(P^X)$ gets larger, we found a dip in PER in the circulant graph case as $\sigma_{\min}(P^X)$ moves from $10^{-31}$ to $10^{-15}$. Though unexpected, this observation is not contradictory to our theory since our theory does not make explicit predictions about the rate of phase transition for \asru/. Across different GAN models, we found that JSD generally approaches perfect \asru/ at a faster rate than MMD in all three graphs, suggesting the use of nonlinear dynamic may be beneficial. Nevertheless, the overall trends for different GANs remain in large part homogeneous. Between Wasserstein and MMD, we observe very little difference in performance, suggesting the regularization effect of NTK is sufficient to control the Lipschitz coefficient of the network. Finally, for the MMD GAN in the matched setting, we found the network is able to achieve perfect \asru/ regardless of the spectral properties of the Markov transition graphs, which confirms our theory that a \smrm/ tends to have simple eigenvalue spectrum suitable for \asru/.

\paragraph{Effect of discriminator reset} As pointed out by \cite{Franceschi2021-gantk}, a discriminator may suffer from residual noise from previous updates and fail to approximate the target divergence measure. We analyze such effects for MMD and JSD as shown in \Figref{fig:discrim_reset}. We observed consistent trends that models whose weights are reset to the initial weights every discriminator loop outperform those without resetting. The effect is more pronounced for JSD GAN than MMD GAN and for smaller $\sigma_{\min}(P^X)$. \paragraph{Effect of generator averaging strategy} The original \wavtovecu/~\cite{Baevski2021-wav2vec-u} directly feeds the text posterior probabilities $O$ into the discriminator, which we refer to as the \emph{``soft input''} approach. Alternatively, we can instead
calculate a weighted average of the gradient form over the samples $y\in\sY^L$
%to do it in a mathematically correct way by averaging the cost weights of the discriminator over such posteriors
as in \Eqref{eq:pde_gen_gd}, which we refer to as the \emph{``outside cost''} approach. The comparison between the two approaches are shown in \Figref{fig:gen_avg}. We observed mixed results: for MMD GANs, the soft-input approach outperforms the outside-cost approach and performs best among the models in the high-$\sigma_{\min}(P^X)$ setting; for JSD GANs, we found that the outside-cost approach performs slightly better than the soft-input approach. Such inconsistencies may be another consequence of the regularization effect predicted by the GANTK. We leave the theoretical explanation as future work.

%% file: related_works.tex
% Who proposed ASR-U?

\cite{Glass2012-unsup-speech} first proposed the challenging task of \asru/ as a key step toward unsupervised speech processing, and framed it as a decipherment problem.
% Who built ASR-U systems? Who achieves the current SOTA?
 \cite{Liu2018-asru} takes on the challenge by developing the first \asru/ system with ground-truth phoneme boundaries and quantized speech features as inputs, by training a GAN to match the speech-generated and real text distributions. \cite{Chen2019-asru} later replaced the ground truth boundaries with unsupervised ones refined iteratively by an HMM, which also incorporates language model information into the system. \cite{Yeh2019-asru} explored the cross entropy loss for matching the generated and real text distribution, but it is prone to mode collapse and needs the help of additional regularization losses such as smoothness weight. More recently, \cite{Baevski2021-wav2vec-u,Liu2022-asru} proposed another GAN-based model using continuous features from the last hidden layer of the \wavtovectwo/~\cite{Baevski2020-wav2vec2} model and additional regularization losses to stabilize training. Their approach achieves ASR error rates comparable to the supervised system on multiple languages, making it the current state-of-the-art system. 
 
To better understand the learning behavior of \asru/ systems, \cite{Lin2022-asru-robustness} analyze the robustness of \wavtovecu/ against empirical distribution mismatch between the speech and text, and found that $N$-gram language model is predictive of the success of \asru/. Inspired by the original framework in \cite{Glass2012-unsup-speech}, \cite{Klejch2022-crosslingual-asru} proposed a decipher-based cross-lingual ASR system by mapping IPA symbols  extracted from a small amount of speech data with unpaired phonetic transcripts in the target language.
% Our analysis of the GAN training for \asru/ is inspired by prior works on the learning theory of GAN.

Our analysis on the sufficient condition of \asru/ is based on previous work on the asymptotic behaviour of GAN objective functions~\cite{Goodfellow2014,Arjovsky17}. Our finite-sample analysis takes inspiration from later work extending the asymptotic analysis to the finite-sample regimes~\cite{Arora2017, Bai2019-approximability-gan}. Such frameworks, however, do not account for the alternate gradient optimization method of GANs and inevitably lead to various inconsistencies between the theory and empirical observations of GAN training~\cite{Franceschi2021-gantk}. 
% They have shown that with an optimal discriminator, the generator is essentially minimizing the divergence measure associated with the GAN objective between the generated and true data distributions. 
% using globally optimal training algorithm
% To better understand the training dynamic of GAN, several works have proved convergence theorems under different restrictive assumptions on the GAN objective~\cite{Mescheder2017, Mroueh2021}, the architecture of the discriminator and/or generator \cite{Domingo2020-mean-field-gan, Balaji2021} and the data distribution~\cite{Mescheder2018}.
Building upon prior works~\cite{Mescheder2017, Mescheder2018, Domingo2020-mean-field-gan, Mroueh2021, Balaji2021}, ~\cite{Franceschi2021-gantk} proposed a unified framework called GANTK based on NTK~\cite{Jacot2018-ntk} to describe the training dynamic of \emph{any} GAN objectives and network architectures. Our analysis on the training dynamic of \asru/ adopts and extends the GANTK framework to handle \emph{discrete, sequential} data such as natural languages.

% Using the NTK theory \textcolor{red}{CITE} showed that under mild regularity conditions, DNN with ReLU activation trained with mean squared error converges to a global optimum. There is a rich literature on the NTK theory since then, and we will not go into details due to space constraint. 

%% file: conclusion.tex
In this paper, we develop a theoretical framework to study the fundamental limits of \asru/ as well as the convergence properties of GAN-based \asru/ algorithms. In doing so, our theory sheds light on the underlying causes of training instability for such algorithms, as well as several new directions for more reliable \asru/ training.

%% file: limitations.tex
Our theory currently assumes that input speech features are quantized into discrete units, as in \cite{Chen2019-asru}, while preserving \emph{all} the linguistic information in the speech. As a result, our theory does not account for the loss of linguistic information during the quantization process, as often occurred in realistic speech datasets. Further, more recent works~\cite{Baevski2021-wav2vec-u, Liu2022-asru} have shown that continuous features, with the help of additional regularization losses, can achieve almost perfect \asru/. Such phenomena is beyond explanations based on our current theory and require generalizing our theory to continuous speech features. Further, our model assumes that sufficiently reliable phoneme boundaries are fed to the \asru/ system, and kept fixed during training. It will be interesting to extend our framework to systems with trainable phoneme boundaries, such as the \wavtovecu/ systems, to better understand its effect on training stability.
% \textcolor{red}{Need quantization and segmentation; necessary and sufficient condition; domain mismatch}

%% file: appendix.tex
\section{Proofs of theoretical results}
\subsection{Learnability of \asru/: a sufficient condition}\label{appendix:sufficient_condition}
\begin{proof}{(Theorem~\ref{thm:asymp_unsup_asr})}
  
  For simplicity, we assume that the eigenvalues of $A$ are real though
  a similar argument applies to complex eigenvalues as well. By
  Assumptions~\ref{assu:eigenvals_of_T} and \ref{assu:init_prob}, it can
  be verified that
  \begin{align*}
    P_{X_{kN}}&=\pi^{\top} A^{k}E\\
    &=\pi^\top U\Lambda^k U^{-1}E,
  \end{align*}
  where $E=\mathbf{1}_{|\sX|^{N-2}}\otimes I_{|\sX|}$, where $\otimes$
  denotes the Kronecker product. Define $c_{jk}=\pi^\top u_{jk}$. Define
  $r_{jk}^\top$ to be the $k^{\textrm{th}}$ row of the $j^{\textrm{th}}$
  block of the matrix $U^{-1}E$, i.e.,
  $UU^{-1}E=\sum_{j=1}^K\sum_{k=1}^{N_j}u_{jk}r_{jk}^\top$.  Define the
  matrix $R_K$ as $R_K=[r_1,\cdots,r_K]$, where $r_j=\sum_{k=1}^{N_j} c_{jk}
  r_{jk}$. Then we have:
  \begin{align*}
    P_{X_{kN}}^\top &= \sum_{j=1}^K\lambda^k_j r_j^\top\\
    P^X &= V_L(\lambda_{1:K})^\top R_K^\top,
  \end{align*}
  where $V_L(\lambda_{1:K})$ is the Vandermonde matrix formed by
  nonzero eigenvalues $\lambda_1,\cdots,\lambda_K$ and with $L$
  columns, $K \geq |\sX|$ by Assumption~\ref{assu:eigenvals_of_T}.
  $R_K$ has full column rank of $K\ge |X|$ by
  Assumption~\ref{assu:init_prob}, therefore it is possible to write
  $R_K=\hat{R}_KL$, where $\hat{R}_K=\hat{r}_1,\ldots,\hat{r}_K]$ is a
    matrix with orthogonal columns, and $L$ is lower-triangular. As a
    result, we have $P^X$ is full rank iff $V_L(\lambda_{1:K})$ has
    full row rank of at least $|\sX|$, which holds by
    Assumption~\ref{assu:eigenvals_of_T}.
  % $\tilde{\pi} = [\tilde{\pi}^1,\cdots, \tilde{\pi}^{|\sX|^{N-2}}], \tilde{\pi}^j:=[\pi_1^j,\cdots,\pi_{|\sX|}^j]$,
\end{proof}

\begin{proof}{(Lemma~\ref{prop:least_singular_val_unsup_asr})}

  Use the Rayleigh-characterization of eigenvalues of the matrix $P^{X\top}P^X$, we have
  \begin{align*}
    &\sigma_{\min}(P^X) \\
    &= \sqrt{\lambda_{\min}(P^{X\top}P^X)}\\
    %&=\sqrt{\min_{w\in \sR^{|\sX|}}\frac{w^\top P^{X\top}P^Xw}{w^\top w}}\\
    %&=\sqrt{\min_{w\in \sR^{|\sX|}}\frac{w^\top R_K V_L V_L^\top R_K^\top w}{w^\top w}} \\
    %&\geq \sqrt{\sum_{l=0}^{L-|\sX|-1}\lambda_{\min}^{2l}
    %  \min_{w\in \sR^{|\sX|}}
    %  \frac{w^\top R_KV_{|\sX|}V_{|\sX|}^\top R_K^\top w}{w^\top w}}\\
    &=\sqrt{\min_{\Vert w\Vert=1}w^\top P^{X\top}P^Xw}\\
    &=\sqrt{\min_{\Vert w\Vert=1}w^\top R_K V_L V_L^\top R_K^\top w}\\
    &\geq \sqrt{\sum_{l=0}^{L-|\sX|-1}\lambda_{\min}^{2l}
      \min_{\Vert w\Vert=1}w^\top R_KV_{|\sX|}V_{|\sX|}^\top R_K^\top w}\\
    &=\sigma_{\min}(P^X_{1:|\sX|})\sqrt{\sum_{l=0}^{L-|\sX|-1}\lambda_{\min}^{2l}},
  \end{align*}
  where $\lambda_{\min}$ is the eigenvalue of $A$ with minimum absolute value,
  and $P^X_{1:|\sX|}$ is the first $\vert\sX\vert$ rows of $P^X$.
  Therefore, to lower bound $\sigma_{\min}(P^X)$, it suffices to lower bound $\sigma_{\min}(P^X_{1:|\sX|})$.
  But note that
  \begin{align*}
    &\sigma_{\min}(P^X_{1:|\sX|})\\  =&\min_{\Vert w\Vert=1}\Vert V_{|\sX|}^TR_K^Tw\Vert\\
    \ge& \sigma_{\min}(V_{|\sX|}^T)\min_{\Vert w\Vert=1}\Vert R_K^Tw\Vert\\
    \ge& \frac{\sigma_{\max}(V_{|\sX|})}{\kappa(V_{|\sX|})}\min_j \Vert \hat{r}_j\Vert\\
    \ge& \frac{|\det(V_{|\sX|})|^{1/|\sX|}}{\kappa(V_{|\sX|})}\min_j \Vert \hat{r}_j\Vert\\
    =& \frac{|\prod_{1\le i<j\le\vert\sX\vert}|\lambda_i-\lambda_j|^{1/|\sX|}}{\kappa(V_{|\sX|})}
    \min_j \Vert \hat{r}_j\Vert\\
    \ge& \frac{\delta_{\min}^{(|\sX|-1)/2|\sX|}}{\kappa(V_{|\sX|})}
    \min_j \Vert \hat{r}_j\Vert
  \end{align*}
  %Further, define the \emph{condition number} of a matrix $A$ as
  %%\begin{align*}
  %    \kappa(A) := \frac{\sigma_{\max}(A)}{\sigma_{\min}(A)},
  %\end{align*}
  %we can lower bound $\sigma_{\min}(P_{1:|\sX|}^X)$ as
  %\begin{align*}
  %    &\sigma_{\min}(P^X_{1:|\sX|})=\frac{\sigma_{\max}(P^X_{1:|\sX|})}{\kappa(P^X_{1:|\sX|})}\\
  %  \geq& \frac{|\det(P^X_{1:|\sX|})|^{1/|\sX|}}{\kappa(P^X_{1:|\sX|})}
  %  =\frac{|\det(V_{|\sX|})\det(R_{|\sX|})|^{1/|\sX|}}{\kappa(P^X_{1:|\sX|})}\\
  %    =& \frac{|\det(R_{|\sX|})|^{1/|\sX|}\prod_{1\le i<j\le\vert\sX\vert}|\lambda_i-\lambda_j|^{1/|\sX|}}{\kappa(P^X_{1:|\sX|})}\\
  %    \geq& \frac{\left(|\det(R_{\vert\sX\vert})|\right)^{1/|\sX|}
  %      \delta_{\min}^{\frac{|\sX|-1}{2|\sX|}}}{\kappa(P^X_{1:|\sX|})},
  %\end{align*}
  where the last equality uses the closed-form formula of the determinant of a square Vandermonde
  matrix,
  %\begin{align*}
  %    \kappa(P_{1:|\sX|}^X) &= \|P_{1:|\sX|}^X\|\|P_{1:|\sX|}^{X+}\|\\
  %    &\leq \|V_{|\xS|}\|\|V_{|\sX|}^+\|\|R_{|\sX|}\|\|R_{|\sX|}^+\| = \kappa(V_{|\sX|})\kappa(R_{|\sX|}),
  %\end{align*}
  and where the behaviour of $\kappa(V_{|\sX|})$, the condition number
  of the Vandermonde matrix, has been studied in depth in
  \cite{Bazan2000}.
  %In order to better characterize the lower bound on
  %$\sigma_{\mathrm{min}}(P_{1:|X|}^X)$, therefore, it is necessary to
  %better characterize the condition number of the projection-scaled
  %eigenvector matrix, $|\det(R_{|\sX|})|/\kappa(V)$, and the minimum gap
  %between the eigenvalues of the transition matrix,
  %$\delta_{\mathrm{min}}$.
\end{proof}

\subsection{Finite-sample learnability of \asru/: matched setup}\label{appendix:proofs_finite_sample}

\paragraph{Theory of \sbp/} The proof of Theorem~\ref{thm:simple_spectrum} makes extensive use of the theory of \sbp/. Therefore, we briefly provide some background on the subject. First, we define the \emph{\sbp/} of a vector $x$ as follows.
\begin{definition}{(Small ball probability)}
Given a fixed vector $x = (x_1, \cdots, x_n)$,  and i.i.d random variables $\xi = (\xi_{i}, \cdots, \xi_{n})$, the small ball probability is defined as
\begin{align*}
    \rho_{\delta}(x) :=\sup_{a\in\sR} \Pr[|\xi^\top x - a|\leq \delta].
\end{align*}
\end{definition}
Intuitively, \sbp/ is the amount of ``\emph{additive} structure'' in $x$: for example, if the coordinates of $x$ are integer multiples of each other and $\xi_i$'s are symmetric Bernoulli variables, the product $\xi^\top x$ tends to have small magnitude as terms cancel out each other very often. Since sparser vectors tend to have less additive structure, \sbp/ can also be used to measure how \emph{sparse} the weights of $x$ are. Another way to look at this is that, if the L2 norm of $x$ is fixed and most of the weight of $x$ is gathered in a few coordinates, the product $\xi^\top x$ has higher variance and is thus less likely to settle in any fixed-length intervals. This is quantitatively captured by the celebrated Offord-Littlewood-Erd\"{o}s (OLE) anti-concentration inequality (and its inverse) for general \subg/ random variables:
\begin{lemma}{\cite{Erdos1945, Rudelson2008, TaoVu2009}}\label{lemma:ole}
Let $\epsilon > 0$ be fixed, let $\delta > 0$, and let $v \in \sR^m$ be a unit vector with
\begin{align*}
    \rho_{\delta}(v) \geq m^{-\frac{1}{2}+\epsilon}.
\end{align*}
Then all but at most $\epsilon m$ of the coefficients of $v$ have magnitude at most $\delta$.
\end{lemma}
Note that here we use a slight generalization of the notion of sparsity called \emph{compressibility} defined as follows.
\begin{definition}{($(\alpha, \delta)$-compressible)}
A vector $v \in \sR^n$ is $(\alpha, \delta)$-compressible if at most $\lfloor\alpha n\rfloor$ of its coefficients have magnitude above $\delta$.
\end{definition}
Note that a sparse vector with a support of size at most $\lfloor\alpha n\rfloor$ is $(\alpha, 0)$-compressible. 

A more generally applicable anti-concentration inequality requires the following definition of generalized arithmetic progression, which is used to quantify the amount of additive structure of a vector.
\begin{definition}{(Generalized arithmetic progression)}
A generalized arithmetic progression (GAP) is a set of the form
\begin{align*}
    Q = \{a^\top w:a\in \sZ^r, |a_i|\leq N_i, 1\leq i\leq r\},
\end{align*}
where $r \geq 0$ is called the \emph{rank} of the GAP and $w_1,\cdots,w_r\in \sR$ are called \emph{generators of the GAP}. Further, the quantity
\begin{align*}
    \mathrm{vol}(Q) := \prod_{i=1}^r (2N_i+1)
\end{align*}
is called the \emph{volume} of the GAP.
\end{definition}

\begin{lemma}{(Continuous inverse Littlewood-Offord theorem, Theorem 2.9 of~\cite{NguyenVu2011})}\label{lemma:generalized_ole}
Let $\epsilon > 0$ be fixed, let $\delta > 0$ and let $v\in\sR^n$ be a unit vector whose \sbp/ $\rho := \rho_{\delta}(v)$ obeys the lower bound
\begin{align*}
    \rho \gg n^{-O(1)}.
\end{align*}
Then there exists a generalized arithmetic progression $Q$ of volume
\begin{align*}
    vol(Q)\leq \max\left(O\left(\frac{1}{\rho \sqrt{\alpha n}}\right), 1\right)
\end{align*}
such that all but at most $\alpha n$ of the coefficients $v_1,\cdots, v_n$ of $v$ lie within $\delta$ of Q. Furthermore, if $r$ denotes the rank of $Q$, then $r = O(1)$ and all the generators $w_1,\cdots,w_r$ of $Q$ have magnitude $O(1)$.
\end{lemma}
 While applicable for any $\rho \gg n^{-\epsilon}$ rather than only those with $\rho_{\delta}(v) \geq n^{-1/2+\epsilon}$ as required by Lemma~\ref{lemma:ole},  Lemma~\ref{lemma:generalized_ole} is \emph{weaker} than Lemma~\ref{lemma:ole} in the sense that rather than showing that the vector is compressible \whp/ and thus covered by the set of compressible vectors, it proves that the vector is covered by a small set \whp/. 

A related notion that is often more convenient for our analysis is the \emph{segmental} \sbp/, which is simply \sbp/ computed on a segment of the vector:
\begin{align*}
    \rho_{\delta, \alpha}(x) = \inf_{I\subseteq \{1,\cdots,n\}:|I|=\lfloor \alpha n\rfloor}\rho_{\delta}(x_I),
\end{align*}
From the definition, it is not hard to see that $\rho_{\delta, \alpha}(x) \geq \rho_{\delta}(x)$.

\paragraph{Eigen-gaps of \smrm/} Armed with tools from the theory of \sbp/, we will establish guarantees of eigenvalue gaps for a \smrm/. First, we shall show that Theorem~\ref{thm:simple_spectrum} implies Corollary~\ref{col:simple_spectrum}. 
\begin{proof}{(Proof of Corollary~\ref{col:simple_spectrum})}
    Using Theorem~\ref{thm:simple_spectrum} and union bound,  the probability that a \smrm/ has at least $|\sX|$ distinct eigenvalues can be bounded as
    \begin{align*}
        & \Pr\left[\min_{1\leq i \leq |\sX|} |\lambda_i-\lambda_{i+1}|\leq |\sX|^{-BN}\right] \leq \\ 
        & |\sX|\max_{i}\Pr\left[|\lambda_i-\lambda_{i+1}|\leq |\sX|^{-BN}\right] \\
        &=  O(|\sX|^{-CN}),
    \end{align*}
    with probability at least $1-O(\exp(-\alpha_0 |\sX|^N))$.
\end{proof}

It turns out that a \smrm/ enjoys various properties analogous to a symmetric matrix. First, we can show that its eigenvalues are real. This can be proved by noting that for a \smrm/ $A_n := D_n^{-1}W_n$ and for any of its eigenvalues $\lambda$ with eigenvector $v$,
\begin{align}\label{eq:smrm_vs_symmetric}
    &D_n^{-1}W_nv = \lambda v\nonumber\\
    \Longleftrightarrow &D_n^{-1/2}W_nD_n^{-1/2}(D_n^{1/2}v) = \lambda D_n^{1/2}v, 
\end{align}
which implies $A_n$ has the same spectrum as $D_n^{-1/2}W_nD_n^{-1/2}$, which is symmetric and thus has a real spectrum. Further, we can prove a variant of Cauchy's interlace theorem for \smrm/.
\begin{lemma}\label{lemma:cauchy_interlacing_thm_for_symmetric_markov}
    Suppose $A_n=D^{-1}_n W_n \in \sR^{n\times n}$ is a \smrm/ with adjacency matrix $W_n$ and eigenvalues $\lambda_1\geq \cdots\geq \lambda_n$ and $A_m=D^{-1}_m W_m$ with adjacency matrix $W_{m-1}$ and eigenvalues $\nu_1\geq \cdots\geq \nu_m, m < n$ is formed by successively deleting $i$-rows and $i$-columns, then
    \begin{align*}
        \lambda_{i} \leq \nu_{i} \leq \lambda_{i+n-m}.
    \end{align*}
\end{lemma}
\begin{proof}
    Using the previous observation in \eqref{eq:smrm_vs_symmetric}, we can apply the standard Cauchy's interlacing theorem on $A_n' := D^{-1/2}_n W_nD^{-1/2}_n$ and $A_m' := D^{-1/2}_m W_mD^{-1/2}_m$, then we have
    \begin{align*}
        \lambda_i(A_n) &= \lambda_i(A_n') \leq \lambda_i(A_m') = \lambda_i(A_m)\\
        &\leq \lambda_{i+n-m}(A_n') = \lambda_{i+n-m}(A_n).
    \end{align*}
\end{proof}

Next, we can show that the eigenvalues of a \smrm/ and its adjacency matrix are simultaneously distributed within the bounded intervals $[-10 n^{\gamma-1}, 10 n^{\gamma-1}]$ and $[-10 n^{\gamma}, 10 n^{\gamma}]$ \whp/. For this and subsequent proofs, we will assume $\gamma'=\gamma> 1/2$.
\begin{lemma}\label{lemma:eigval_dist_symmetric_markov}
Let $A_n = D_n^{-1}W_n$ be a \smrm/ with adjacency matrix $W_n$ and properties defined in Theorem~\ref{thm:simple_spectrum}, then we have with probability at least $1-O(\exp(-\alpha_0 n))$,
\begin{align*}
    \lambda_i(A_n) &\in [-10n^{\gamma-1}, 10n^{\gamma-1}],\\
    \lambda_i(W_n) &\in [-10n^{\gamma}, 10n^{\gamma}],
\end{align*}
%where $\gamma'=\max\{\gamma, 1/2\}$,
for any $1\leq i \leq n$ and some $\alpha_0 > 0$.
\end{lemma}
\begin{proof}
     First, by definition, we can let $W_n = F_n + X_n$, where $F_n$ is a deterministic matrix with eigenvalues of order $n^{\gamma}$ and $X_n$ is a symmetric matrix whose elements are independent zero-mean unit-variance \subg/ random variables. Using standard results from random matrix theory~\cite{Anderson2009-rmt}, we have
    \begin{align*}
        \{\lambda_1(X_n), \cdots, \lambda_n(X_n)\} \subset [-10n^{\gamma-1}, 10n^{\gamma-1}],
    \end{align*}
    with probability at least $1-O(\exp(-\alpha_0 n))$. Therefore, Weyl's matrix perturbation inequality then ensures that
    \begin{align*}
         \{\lambda_1(W_n), \cdots, \lambda_n(W_n)\} \in [-10n^{\gamma}, 10n^{\gamma}],
    \end{align*}
    with probability at least $1-O(\exp(-\alpha_1 n))$. Suppose this event occurs and use Lemma~\ref{lemma:cauchy_interlacing_thm_for_symmetric_markov} and the variational characterization of eigenvalues, we have
\begin{align*}
    \lambda_i(A_n) &= \min_{V_{i-1}}\max_{v \in V_{i-1}^{\perp}\atop\|v\|_2=1}v^\top D_n^{-1/2}W_nD_n^{-1/2}v\\
    &=\min_{V_{i-1}}\max_{v \in V_{i-1}^{\perp}\atop v^\top D v=1}v^{\top}W_nv\\
    &= \min_{V_{i-1}}\max_{v \in V_{i-1}^{\perp}}\frac{v^{\top}W_nv}{v^\top D_n v},
\end{align*}
where $V_{i-1}$ is a subspace of dimension $i-1$.
Combining the two results, we have with probability at least $1-O(\exp(-\alpha_1 n))$,
%Combining the two results, we have with probability at least $1-O(\exp(-\max\{\alpha_0, \beta_0\} n))$,
%\begin{align*}
%    &\max_{v \in V_{i-1}^{\perp}}\frac{v^{\top}Wv}{v^\top D v} = \max_{v \in V_{i-1}^{\perp}\atop \|v\|_2=1}\frac{v^{\top}Wv}{v^\top D v}\\
%    &\geq \frac{\min_{v:\|v\|=1}v^{\top}Wv}{\max_{v:\|v\|=1}v^{\top}Dv} = \frac{\lambda_n(W)}{\max_{i}d_{ii}} = -O(n^{\gamma-1}),
%\end{align*}
%and
\begin{align*}
  \max_{v \in V_{i-1}^{\perp}}\left\vert\frac{v^{\top}W_nv}{v^\top D_n v}\right\vert
  &\leq \frac{\max_{v:\|v\|=1}\vert v^{\top}W_nv\vert}{\min_{v:\|v\|=1}\vert v^{\top}D_nv\vert} \\
  &= \frac{\lambda_1(W)}{\min_{i}\vert d_{ii}\vert}
\end{align*}
Recall that $d_{ii}=\sum_{j=1}^n w_{ij} =\sum_{j=1}^n
 (f_{ij}+x_{ij})$, where $w_{ij}$, $f_{ij}$, and $x_{ij}$ are the
$(i,j)^{\textrm{th}}$ elements of $W_n,F_n$, and $X_n$ respectively.
Since $A_n=D_n^{-1}W_n$ is a Markov matrix we assume that $f_{ij}$ and
the distribution of $x_{ij}$ are selected to guarantee that $w_{ij}\ge
0$, e.g., it must be true that $f_{ij}\ge 0$.  We also know that
$x_{ij}$ is a zero-mean unit-variance sub-Gaussian random variable,
therefore
\begin{align*}
  \Pr\left\{w_{ij}<\delta\right\} &=
  \Pr\left\{x_{ij}<-f_{ij}+\delta\right\} \\
  &\le 2\exp\left(-\frac{1}{2}(f_{ij}-\delta)^2\right)\\
  \Pr\left\{d_{ii}<n\delta\right\} &=
  \Pr\left\{\sum_{j=1}^n w_{ij}<n\delta\right\} \\
  & \le 2\exp\left(-\alpha_2 n\right)\\
\end{align*}
where $\alpha_2=-\frac{1}{2}(\bar{f}_{i}-\delta)^2$, and $\bar{f}_i=\frac{1}{n}\sum_jf_{ij}$.
Therefore, with probability at least $1-O(\exp(-\alpha_0 n))$ where $\alpha_0=\alpha_1+\alpha_2$, 
%which implies
%\begin{align}\label{eq:eigval_range_symmetric_markov}
%    \lambda_{i}(A) \in [-10n^{\gamma-1}, 10n^{\gamma-1}],\,1\leq i\leq n
%\end{align}
%\begin{align*}
%    \max_{v \in V_{i-1}^{\perp}}\frac{v^{\top}Wv}{v^\top D v} &\leq \frac{\max_{v:\|v\|=1}v^{\top}Wv}{\min_{v:\|v\|=1}v^{\top}Dv} \\
%    &= \frac{\lambda_1(W)}{\min_{i}d_{ii}} = O(n^{\gamma-1}),
%\end{align*}
%which implies
\begin{align}\label{eq:eigval_range_symmetric_markov}
    \lambda_{i}(A_n) \in [-10n^{\gamma-1}, 10n^{\gamma-1}],\,1\leq i\leq n
\end{align}
%with probability at least $1-O(\exp(-\alpha_0 n))$. 
\end{proof}

\paragraph{Remark.} Lemma~\ref{lemma:eigval_dist_symmetric_markov} ensures that for any \smrm/ $A_n=D_n^{-1}W_n$ with properties defined in Theorem~\ref{thm:simple_spectrum}, we can focus our attention on any eigenvector $v$ whose eigenvalue is no greater than $O(n^{\gamma-1})$ and whose $\|W_nv\|_2$ is of order $n^{\gamma}$ \whp/. Therefore, we will assume such conditions in later proofs.
% Another property of \smrm/ analogous to symmetric random matrices concern the (in)-compressibility of its eigenvectors.

Using Lemmas~\ref{lemma:cauchy_interlacing_thm_for_symmetric_markov}-\ref{lemma:eigval_dist_symmetric_markov}, we can reduce Theorem~\ref{thm:simple_spectrum} to the following statement on \sbp/ of the \emph{eigenvectors} of $X_n$, analogous to the arguments for symmetric random matrices in \cite{NguyenTaoVu2015}.

\begin{lemma}\label{lemma:smrm_eigenval_gap_to_eigenvec_sbp}
Let $A_n = D_n^{-1}W_n\in \sR^{n\times n}$ be a \smrm/ with adjacency matrix $W_n$.
Let $\lambda_i(A_n)$
%be the $i$-th eigenvalue of the matrix $A$
and $w=[u^\top, b]^\top\in \sR^n$ be the $i$-th eigenvalue and eigenvector of the matrix $A_n$, respectively, where $u\in\sR^{n-1}$ and $b\in\sR$. Then we have
\begin{multline*}
    \Pr[|\lambda_i(A_{n})-\lambda_{i+1}(A_{n})|\leq \delta]\leq\\
    n\Pr[\rho_{\delta n^{\gamma+1}}(v) \geq c_0n^{\gamma+1}\delta] + c_0 n^{\gamma+2}\delta \\
    + O(\exp(-\alpha_0 n)), % \Pr\left[\frac{|v^\top X|}{\max_{1\leq i\leq n}d_{ii}}\leq \frac{\delta}{b} \right].
    %n\rho_{c_1\delta n^{3/2}}(v) + O(\exp(-\alpha_0 n)),
    % \Pr\left[\frac{|v^\top X|}{\max_{1\leq i\leq n}d_{ii}}\leq \frac{\delta}{b} \right].
\end{multline*}
for some $c_0,\alpha_0 > 0$.
\end{lemma}
\begin{proof}
  %By definition, we have
  Let $W_{n-1}$ and $D_{n-1}$ be the $(n-1)$-dimensional minors of $W_n$ and $D_n$, respectively, then
    \begin{align*}
        \begin{bmatrix}
    W_{n-1} & w_n \\
    w_n^\top & w_{nn}
    \end{bmatrix} \begin{bmatrix}
    u \\
    b
    \end{bmatrix} = \lambda \begin{bmatrix}
    D_{n-1} & \mathbf{0}_n\\
    \mathbf{0}_n^\top & d_{nn}
    \end{bmatrix}\begin{bmatrix}
    u \\
    b
    \end{bmatrix},
    \end{align*}
    where $w_n$ is the last column of $W_n$.
    Let $v$ be the $i$-th eigenvector of matrix $A_{n-1} := D_{n-1}^{-1}W_{n-1}$, we have 
    \begin{align*}
        &v^\top W_{n-1}u + v^\top Wb = \lambda_{i}(A_n) v^\top D_{n-1}u\\
        \Longrightarrow & |(\lambda_{i}(X_{n-1})-\lambda_i(X_n))|\max_{1\leq i\leq n}d_{ii} \geq \\
        &|(\lambda_{i}(A_{n-1})-\lambda_i(A_n)) v^\top D_{n-1} u| = |v^\top w_n b|.
    \end{align*}
    Therefore,
    \begin{align*}
        &\Pr[|\lambda_i(A_{n})-\lambda_{i}(A_{n-1})|\leq \delta] \\
        \leq& \Pr\left[\frac{|v^\top w_n|}{\max_{1\leq i\leq n}d_{ii}}\leq \frac{\delta}{b} \right]. 
    \end{align*}
    By Lemma~\ref{lemma:cauchy_interlacing_thm_for_symmetric_markov}, $\lambda_{i+1}(A_n) \leq \lambda_{i}(A_{n-1}) \leq \lambda_{i}(A_n)$ and we have
    \begin{multline*}
        \Pr[|\lambda_i(A_{n})-\lambda_{i+1}(A_{n})|\leq \delta] \leq\\ \Pr[|\lambda_i(A_{n-1})-\lambda_{i}(A_{n})|\leq \delta] \le \\
         \Pr\left[\frac{|v^\top w_n|}{\max_{1\leq i\leq n}d_{ii}}\leq \frac{\delta}{b} \right].
    \end{multline*}
    $d_{ii}$ is typically $O(n)$, but we have been unable to prove
    that it is necessarily $O(n)$.  Consider that
    $w_{ij}=f_{ij}+x_{ij}$, where $F_n$ is a symmetric matrix with
    eigenvalues $\lambda_i(F_n)=O(n^\gamma)$, therefore
    \begin{align*}
      \sum_{j=1}^n f_{ij} &=
      (F_n\mathbf{1}_n)_i \le \|F_n\mathbf{1}_n\|_2 = \|F_n\|_1 \\
      &\leq n^{1/2}\|F_n\|_2 = O\left(n^{\gamma+\frac{1}{2}}\right).
    \end{align*}
    $W_n=F_n+X_n$, therefore
    \begin{align*}
      &\Pr\left\{d_{ii}\ne O\left(n^{\gamma+\frac{1}{2}}\right)\right\}\\
      &\le\Pr\left\{\sum_{j=1}^nx_{ij}>\sum_{j=1}^nf_{ij}-n\delta\right\}\\
      &\le O(\exp(-\alpha_2 n))
    \end{align*}
    Now, by the law of total probability,
    \begin{align*}
        &\Pr\left[\frac{|v^\top w_n|}{\max_{1\leq i\leq n}d_{ii}}\leq \frac{\delta}{b} \right] \\
      \leq& \Pr\left[\frac{|v^\top w_n|}{\max_{1\leq i\leq n}d_{ii}}\leq \frac{\delta}{b},
        \max_{1\le i\le n}d_{ii} \leq O(n^{\gamma+\frac{1}{2}})\right]\\
        %d_{ii} \leq O(n)\right]\\
        & + \Pr\left[\max_{1\le i\le n}d_{ii}\ne O\left(n^{\gamma+\frac{1}{2}}\right)\right]\\
        %& + \Pr\left[d_{ii} > n + o(n)\right]\\
      \leq& \Pr\left[|v^\top w_n|
        = O\left(\frac{\delta n^{\gamma+\frac{1}{2}}}{b}\right)\right] + O(\exp(-\alpha_2 n)).
        %\leq& \Pr\left[|v^\top W|\leq \frac{\delta (n+o(n))}{b}\right] + O(\exp(-\alpha_0 n)).
    \end{align*}
    By symmetry, we can choose any row and the corresponding column to split the matrix and derive inequality of the same form. Further, suppose for some $b_1 > 0$, with probability at least $1-\exp(-c_1 n)$, there are at least $n_T$ coordinates of $w$ that are at least $b_1$ and suppose we choose the split index $J$ uniformly at random. Let the $J$-th column of $W_{n}$ be $W$ and the $J$-th coefficient of the eigenvector of $W_n$ be $w_J$, then we have
    \begin{align*}
        & \Pr[|\lambda_i(A_{n})-\lambda_{i+1}(A_{n})|\leq \delta]\\
      \leq & \Pr\left[|v^\top W|\ne O\left(\frac{\delta n^{\gamma+\frac{1}{2}}}{w_J} \right)|N_b \geq n_b\right] \\
        & + O(\exp(-c_1n)) + O(\exp(-\alpha_2 n))\\
      \leq & \frac{n}{n_T}\Pr\left[|v^\top W|\ne O\left(\frac{\delta n^{\gamma+\frac{1}{2}}}{b_1}\right)|N_b \geq n_b
        \right] \\
        & + O(\exp(-c_1 n)) + O(\exp(-\alpha_2 n)),
        %\leq & \Pr\left[|v^\top w_n|\leq \frac{\delta (n+o(n))}{w_J}\right|w_J \geq b_1] \\
        %& + O(\exp(-\alpha_0 n))\\
        %\leq & \frac{n}{n_T}\Pr\left[|v^\top w_n|\leq \frac{\delta (n+o(n))}{b_1}\right]\\
        %& + O(\exp(-c_1 n)) + O(\exp(-\alpha_0 n)),
    %Notice that by symmetry, we can choose any row $j$ and its corresponding column to split the matrix and derive an inequality of the same form by redefining $W$ and $b$ accordingly. Further, suppose for some $b_1 > 0$, with probability at least $1-\exp(-c_1 n)$, the number of coordinates of $w$ that are at least $b_1$, $N_{b}$, is at least $n_b$ and suppose we choose the split index $J$ uniformly at random, we have $b=w_J$ and
    %\begin{align*}
    %    & \Pr[|\lambda_i(X_{n})-\lambda_{i+1}(X_{n})|\leq \delta]\\
    %    \leq & \Pr\left[|v^\top W|\leq \frac{\delta (n+o(n))}{w_J}\right] + O(\exp(-\alpha_0 n))\\
    %    \leq & \frac{n}{n_b}\Pr\left[|v^\top W|\leq \frac{\delta (n+o(n))}{b_1}| N_b \geq n_b\right]\\
    %    & + O(\exp(-\min\{c_1, \alpha_0\} n)),
    \end{align*}
    where the second inequality can be proved as follows. Define 
    \begin{align*}
        \gE &= \left\{N_b \geq n_b \right\},\\
        \gF &= \left\{w_J \geq b_1\right\},\\
        \gG &= \left\{|v^\top W|\ne O\left(\frac{\delta n^{\gamma+1/2}}{w_J}\right)\right\},\\
        \gH &= \left\{|v^\top W|\ne O\left(\frac{\delta n^{\gamma+1/2}}{b_1}\right)\right\}.
    \end{align*}
    Then use the above definitions and the fact that $\gF$ and $\gG$ are conditionally independent given $N_b$, we have
    \begin{align*}
        & \Pr\left[|v^\top W| \ne O\left(\frac{\delta n^{\gamma+\frac{1}{2}}}{b_1}\right)|N_b \geq n_b\right]\\
        = & \Pr(\gH|\gE) \geq \Pr(\gF\cap\gG|\gE) \geq \frac{n_T}{n} \Pr(\gG|
        \gE) \\
        = & \frac{n_T}{n}\Pr\left[|v^\top W|\ne O\left(\frac{\delta n^{\gamma+1/2}}{w_J}\right)|N_b \geq n_b\right].
        %& \Pr\left[|v^\top W| \leq \frac{\delta(n+o(n))}{b_1}\right] \\
        %\geq & \Pr\left[|v^\top W| \leq \frac{\delta(n+o(n))}{w_J}, w_J \geq b_1\right]\\
        %\geq & \frac{n_T}{n} \Pr\left[|v^\top W|\leq \frac{\delta(n+o(n))}{w_J}|w_J \geq b_1\right].
    \end{align*}
        %& \Pr\left[|v^\top W| \leq \frac{\delta(n+o(n))}{b_1}|N_b \geq n_b\right]
        %=: \Pr(\gH|\gE)\\
        %\geq & \Pr(\gF \cap \gG|\gE) \geq \frac{n_b}{n}\Pr(\gG|\gE) \\
        %= & \frac{n_b}{n}\Pr\left[|v^\top W|\leq \frac{\delta (n+o(n))}{w_J}|N_b \geq n_b\right].
    %\end{align*}
    Further, to remove the dependency on $N_b$, notice that
    \begin{align*}
        \Pr(\gH|\gE) &\leq \frac{\Pr(\gH)}{\Pr(\gE)}
        = \Pr(\gH) + O(\exp(-c_1 n)).
    \end{align*}
    Next, by the pigeonhole principle, at least one coordinate of the unit eigenvector $w$ is at least $n^{-1/2}$, and thus we can let $c_1 = \infty$, $n_b=1$ and $b_1=n^{-1/2}$ and arrive at
    \begin{align}
      & \Pr\left[|\lambda_i(A_{n})-\lambda_{i+1}(A_{n})|\leq \delta\right]\nonumber\\
      \leq & n\Pr\left[|v^\top W|\ne O\left(\delta n^{\gamma+1}\right)\right]+ O(e^{-\alpha_0 n})\nonumber\\
      \leq & n\rho_{\delta O(1)n^{\gamma+1}}(v) + O(\exp(-\alpha_0 n)),\label{eq:lemma6}
    \end{align}
    where $\alpha_0=c_1+\alpha_2$. Finally, recall the definition of \sbp/, we have
    \begin{multline*}
        \Pr\left[|v^\top W|\leq \delta \right] \leq \Pr\left[|v^\top W|\leq \delta | \rho_{\delta}(v) \leq \epsilon\right] \\ 
        + \Pr[\rho_\delta(v) > \epsilon] \\
        \leq  \Pr[\rho_\delta(v) > \epsilon] + \epsilon,
    \end{multline*}
    and thus applying this inequality with $\delta := c_0 \delta n^{\gamma+1}$ on \Eqref{eq:lemma6} yields the result.
\end{proof}
\paragraph{Remark.} We can sharpen the bound in Lemma~\ref{lemma:smrm_eigenval_gap_to_eigenvec_sbp} by extending the delocalization theorem for a symmetric Wigner matrix (see Theorem 4.2 of ~\cite{NguyenTaoVu2015}) to a \smrm/ and using it to choose a larger $n_b$ in the proof. This will be left as future work. 

With the help of Lemma~\ref{lemma:smrm_eigenval_gap_to_eigenvec_sbp}, we can reduce Theorem~\ref{thm:simple_spectrum} to the following theorem.
\begin{theorem}\label{thm:sbp_random_eigenvec_symmetric_markov}
Let $A_n \in \sR^{n\times n}$ be a \smrm/ matrix and $v$ be an eigenvector with eigenvalue $\lambda = O(n^{\gamma-1})$, then for any fixed $C > 0$, there exists some $B > \max\{4\gamma C + 3\gamma, 4\gamma + 1\}$ such that
\begin{align*}
    \rho_{n^{-B}}(v) \leq n^{-C},
\end{align*}
with probability at least $1-O(\exp(-\alpha_0n))$ for some $\alpha_0$ depending on $B$.
\end{theorem}

Similar to the proof for the perturbed symmetric matrices in \cite{NguyenTaoVu2015}, we reduce Theorem~\ref{thm:sbp_random_eigenvec_symmetric_markov} to the following.
\begin{theorem}\label{thm:sbp_random_eigenvec_symmetric_markov_reduction}
Let $v$ be the eigenvector and $B$ be the constant defined in Theorem~\ref{thm:sbp_random_eigenvec_symmetric_markov}. Then for any $n^{-B} \leq \delta \leq n^{-B/2}$, we have with probability $O(\exp(-\alpha_0 n))$,
\begin{align}\label{eq:thm_reduction_symmetric_markov}
    n^{-C} \leq \rho_{n^{\gamma}\delta}(v) \leq n^{0.49}\rho_{\delta}(v). 
\end{align}
\end{theorem}

To show that Theorem~\ref{thm:sbp_random_eigenvec_symmetric_markov_reduction} implies Theorem~\ref{thm:sbp_random_eigenvec_symmetric_markov}, we prove the contrapositive of the statement, that is, if $\rho_{n^{-B}}(v) > n^{-C}$, then there exists $n^{-B}\leq \delta \leq n^{-B/2}$ such that \eqref{eq:thm_reduction_symmetric_markov} holds with probability at least $1-O(\exp(-\alpha_0 n))$. To construct such $\delta$, let
\begin{align*}
    \delta_0 &:= n^{-B}\\
    \delta_{j+1} &:= n^{\gamma}\delta_{j}, 
\end{align*}
for $j=0,\cdots,J-1$ with $J=\lfloor B / 2\gamma\rfloor$. By construction, we have
\begin{align*}
    &n^{-B}= \delta_0 \leq \delta_j \leq \delta_J \leq n^{-B/2}\\
    &\rho_{\delta_{j}}(v) \geq \rho_{\delta_0}(v) \geq n^{-C}.
\end{align*}
Suppose \eqref{eq:thm_reduction_symmetric_markov} does not hold for any $\delta:=\delta_j$, or otherwise the result follows, we have
\begin{align*}
    \rho_{\delta_J}(v) \geq n^{0.49 J}\rho_{n^{-B}}(v) \geq n^{0.49\lfloor B / 2\gamma\rfloor -C} > 1,
\end{align*}
if $B \geq 4\gamma C + 3\gamma$, which contradicts the fact that $\rho_{\delta_J}(v)\leq 1$. As a result, there has to exist some $j$ such that \eqref{eq:thm_reduction_symmetric_markov} holds.

Again similar to the perturbed symmetric matrix case in \cite{NguyenTaoVu2015}, we divide the proof of Theorem~\ref{thm:sbp_random_eigenvec_symmetric_markov_reduction} into the compressible case and the non-compressible case. For the compressible case, we first prove the following lemma.
\begin{lemma}\label{lemma:compressible_case_symmetric_markov}
Suppose $v$ is an eigenvector of a \smrm/ $A_n := D_n^{-1}W_n$ with adjacency matrix $W_n$ and the same properties defined in Theorem~\ref{thm:simple_spectrum}, and suppose there exists $\delta \in [n^{-B}, n^{-B/2}]$ such that $\rho_{\delta, \alpha}(v) \geq (\alpha n)^{-1/2+\epsilon}$, we have with probability $O(\exp(-\alpha_0 n))$,
\begin{align*}
    n^{-C} \leq \rho_{n^{\gamma}\delta}(v) \leq n^{0.49}\rho_{\delta}(v). 
\end{align*}
\end{lemma}
\begin{proof}
Using concentration inequalities, we have with probability at least $1-O(\exp(-\alpha_2 n))$ for some $\alpha_2 > 0$,
\begin{align}\label{eq:norm_factor_range_symmetric_markov}
  d_{ii} = O\left(n^{\gamma+\frac{1}{2}}\right),~1\le i\le n
    %d_{ii} = n + o(n), 1\leq i\leq n.
\end{align}
Further, since $\rho_{\delta, \alpha}(v) \geq (\alpha n)^{-1/2+\epsilon}$, by Lemma~\ref{lemma:ole}, we have $v$ is $(O(\alpha), \delta)$ compressible, and thus there exists $I$ of of size $O(\alpha n)$ such that $v_i > \delta$ only if $i \in I$. Without loss of generality, let $I=\{n-k,\cdots,n\}$ for $k=O(\alpha n)$ and $\mathbb{E}[A_{ij}]=1$. Further, split $v = [v'^\top, v''^\top]^\top$, then by definition of eigenvalues and eigenvectors,
\begin{align*}
    \begin{bmatrix}
    W_{n-k} & F \\
    F^\top & W_{k}
    \end{bmatrix} \begin{bmatrix}
    v' \\
    v''
    \end{bmatrix} = \lambda \begin{bmatrix}
    D_{n-k} & \mathbf{0} \\
    \mathbf{0}^\top & D_k
    \end{bmatrix}\begin{bmatrix}
    v' \\
    v''
    \end{bmatrix}.
\end{align*}
Reading off the first line of the matrix equation, we have
\begin{align*}
    \|Fv''\|_2 &= \|(W_{n-k}-\lambda D_{n-k}) v'\|_2\\
    &\leq \|W_{n-k}v'\|_2 + \|\lambda D_{n-k} v'\|_2.
\end{align*}
Notice that assuming \eqref{eq:norm_factor_range_symmetric_markov} and \eqref{eq:eigval_range_symmetric_markov} occur, we have
that all elements $v_i'$ of $v'$ have $|v_i'|<\delta$, therefore $\Vert v'\Vert_2\le\delta n^{-1/2}$, therefore
\begin{align*}
    \|W_{n-k}v'\|_2 &\leq \delta n^{1/2}\max_{v:\|v\|_2=1}\|Wv\|_2 \\
    &= O(n^{-B/2+1/2+\gamma})
\end{align*}
Furthermore, if we assume that \Eqref{eq:eigval_range_symmetric_markov} and \Eqref{eq:norm_factor_range_symmetric_markov} occur, then
\begin{align*}
    %\|\lambda D_{n-k}v'\|_2 &= O(n^{\gamma-1}\cdot n \cdot \delta n^{1/2})\\
    %&= O(n^{-B/2+1/2+\gamma}). 
    \|\lambda D_{n-k}v'\|_2 &= O(n^{\gamma-1}\cdot n^{\gamma+1/2} \cdot \delta n^{1/2})\\
    &= O(n^{-B/2+2\gamma}). 
\end{align*}
Thus, using the fact that $B \geq 4\gamma + 1$,
\begin{align*}
    \|Fv''\|_2 &= O(n^{-B/2+2\gamma})=O(n^{-1/2}).
\end{align*}
On the other hand, using a standard epsilon-net argument, with probability at least $1-O(\exp(-\alpha_3 n))$,
\begin{align*}
  \inf_{w\in \sR^{k}:\|w\|=1}\|Fw\|_2
  %\gg (\alpha n)^{1/2}
  \geq n^{-1/2}.
\end{align*}
Now, define the events
\begin{align*}
    \gE &:= \{v \text{ is an eigenvector of }A\}\\ 
    \gE_{\alpha, \delta} &:= \{v \text{ is }(O(\alpha), \delta)\text{-compressible}\}\\
    \gE_{I} &:= \{\|W_{I^c, I}v_I\|_2 \gg O(n^{-1/2})\},
\end{align*}
then by the previous discussion, we have
\begin{align*}
    &\Pr(\gE_I|\gE\cap\gE_{\alpha, \delta}) = O(\exp(-\alpha_2 n))\\
    &\Pr(\gE_I^c|\gE) = O(\exp(-\alpha_3 n)).
\end{align*}
Note that to prove the lemma, it suffices to show that the eigenvector $v$ is not $(O(\alpha), \delta)$-compressible \whp/, or $\Pr(\gE_{\alpha, \delta}|\gE)$ is small, since that will lead to $\rho_{\delta, \alpha}(v) < (\alpha n)^{-1/2+\epsilon}$ \whp/ and thus a contradiction \whp/. Indeed, we have
\begin{align*}
    \Pr(\gE_{\alpha, \delta}|\gE) &\leq \Pr(\gE_{\alpha, \delta}\cap\gE_{I}|\gE) + \Pr(\gE_{\alpha,\delta}\cap\gE_{I}^c|\gE)\\
    &\leq \Pr(\gE_{I}|\gE\cap\gE_{\alpha,\delta}) + \Pr(\gE_{I}^c|\gE)\\
    &= O(\exp(-\alpha_0 n))
\end{align*}
for some $\alpha_0 > 0$.
\end{proof}

%\textcolor{red}{TODO Check this part from now on}
For the incompressible case, we apply the continuous inverse Offord-Littlewood theorem to discretize the set of eigenvectors, and prove the following result analogous to the symmetric case in \cite{NguyenVu2011}.
\begin{lemma}\label{lemma:incompressible_case_symmetric_markov}
Suppose $v$ is an eigenvector of a \smrm/ $A_n := D_n^{-1}W_n$ with adjacency matrix $W_n$ and the same properties defined in Theorem~\ref{thm:simple_spectrum}, and suppose there exists $\delta \in [n^{-B}, n^{-B/2}]$ such that $q:=\rho_{\delta, \alpha}(v) < (\alpha n)^{-1/2+\epsilon}$, we have with probability $O(\exp(-\alpha_0 n))$,
\begin{align*}
    n^{-C} \leq \rho_{n^{\gamma}\delta}(v) \leq n^{0.49}\rho_{\delta}(v). 
\end{align*}
\end{lemma}

To prove this result, we need the following useful lemmas.
\begin{lemma}\label{lemma:discretize_eigenvec_symmetric_markov}
  For any eigenvector-eigenvalue pair $(v, \lambda)$ and $\alpha > 0$ with $|\lambda| = O(n^{\gamma-1})$, suppose $n^{-C}<\rho_{\delta,\alpha}(v) =: q \leq (\alpha n)^{-1/2+\epsilon}$, then
  with probability at least $1-O(\exp(-\alpha_0 n))$
  there exists a subset
  $\gN$ of $\sR^n\times \sR$ of size $O(n^{-n/2+O(\alpha n)}q^{-n+O(\alpha n)})$ such that, there exists $(\tilde{v}, \tilde{\lambda}) \in \gN$ with the properties:
    %$\gN$ of $\sR^n\times \sR^k \times \sR$ of size $O(n^{-n/2+O(\alpha n)}q^{-n})$ such that, there exists $(\tilde{v}, w', \tilde{\lambda}) \in \gN$ with the properties:
\begin{enumerate}
\item $|v_j-\tilde{v}_j|\leq \delta$ for $1\leq j\leq n$;
    %MH: I don't think the following one is necessary!
    %\item $|v_j'-w_j'|\leq n^{-\gamma-1}\delta$ for $1\leq j\leq \lfloor \alpha n\rfloor$;
\item $|\lambda-\tilde{\lambda}|\leq n^{\gamma}\delta$.
    %MH: I think this one can be relaxed a lot, as shown above
    %\item $|\lambda-\tilde{\lambda}|\leq n^{-\gamma-2}\delta$.
\end{enumerate}
\end{lemma}
\begin{proof}
  Split $\{1,\cdots,n\}$ into sets of length differing by at most $1$, $I_1,\cdots,I_m$, $m=\left\lfloor \frac{1}{\alpha}\right\rfloor+1$, then we have
  the length of each set is greater than or equal to $\lfloor\alpha n\rfloor$, and its small ball probability is
    \begin{align*}
        \rho_{\delta}(v_{I_i})\geq \rho_{\delta,\alpha}(v) = q, 1\leq i \leq m.
        %\rho_{\delta}(v_{I_i})\geq \rho_{\delta,\alpha}(v) \geq \rho_{\delta}(v) > q, 1\leq i \leq m.
    \end{align*}
    Therefore, since $q\le(\alpha n)^{-\frac{1}{2}+\epsilon}$ and $n^{-C}<q$, 
    %Applying Lemma~\ref{lemma:generalized_ole} and the assumption that $q\leq (\alpha n)^{-1/2+\epsilon}$,
    there exists a GAP
    \begin{align*}
      Q_i = \left\{\sum_{j=1}^{r_i}a_{ij}w_{ij}:
      \begin{array}{l}
        a_j\in \mathbb{Z},\\
        |a_{ij}|\leq N_{ij}, \\
        1\leq j \leq r_i\end{array}
        \right\}
    \end{align*}
    such that
    \begin{align*}
        \sup_{j\in I_i \setminus S}\inf_{\tilde{v}_j\in Q_i}|v_j-\tilde{v}_j|\leq \delta,
    \end{align*}
    with volume
    \begin{align*}
        \text{vol}(Q_i) \le O((\alpha n)^{-1/2+\epsilon}/q), 1\leq i\leq m,
    \end{align*}
    for all except at most $O(\alpha^2 n)$ indices from some exceptional set $S$. Further, for each $Q_i$, we can quantize its generators $w_{i1},\cdots,w_{ir_i}$ to the closest multiple of $q\delta$, $\tilde{w}_{i1},\cdots,\tilde{w}_{ir_i}$. This introduces an additional approximation error of at most
    \begin{align*}
        &\left|\sum_{j=1}^{r_i}a_{ij}w_{ij}-\sum_{j=1}^{r_i}a_{ij}\tilde{w}_{ij}\right|\\
        \leq & \text{vol}(Q_i)\cdot q\delta \leq (\alpha n)^{-1/2+\epsilon}/q\cdot q\delta \\
        = & (\alpha n)^{-1/2+\epsilon}\delta = O(\delta).
    \end{align*}
    Next, for the coefficients from the exceptional set $S$, we also round them to the closest multiple of $q\delta$ and let the set of such values be $R$, which ensures that 
    \begin{align*}
        \sup_{j\in S}\inf_{v'\in R}|v_j-v'|= O(\delta).
    \end{align*}
    Therefore, for fixed generators $w_{ij}$'s and a given $S$, we can construct a finite set of vectors
    \begin{align*}
      \{\tilde{v}:\tilde{v}_j \in \cup_{i=1}^m Q_i,\,\forall j\not\in S\text{ and }v_j'\in R,\,\forall j\in S\}
      %\{v':v_j' \in \cup_{i=1}^m Q_i,\,\forall j\not\in S\text{ and }v_j'\in R,\,\forall j\in S\}
    \end{align*}
    of size at most
    \begin{align*}
        &\left(m\sup_i \text{vol}(Q_i)\right)^{n-|S|}|R|^{|S|}\\
        \leq & O\left(\frac{1}{\alpha}\frac{(\alpha n)^{-1/2+\epsilon}}{q} \right)^n \cdot O((1/q\delta)^{O(\alpha n)})\\
        \leq & O\left(n^{-\frac{n}{2}+\epsilon n}q^{-n+O(\alpha n)}\right)O\left(n^{B\alpha n}\right)\\
        = & O(n^{-n/2+O(\alpha n)}q^{-n+O(\alpha n)}),\\
        = & O(n^{-n/2+O(\alpha n)}q^{-n}),
    \end{align*}
    that approximates $v$ within $O(\delta)$ for every coefficients.
    The third line uses $\delta>n^{-B}$ and $\alpha=O(1)$; the fourth line assumes $\epsilon=O(\alpha)$.
    %Note that we use the fact that $q, \delta \geq n^{-O(1)}$.
    Further, if we allow the generators to be variable and assume $S$ to be unknown, the quantization mentioned previously and the crude bound of the number of possible $S$ by $2^n$ enlarges the set of vectors by a factor of
    \begin{align*}
        &O\left((1/q\delta)^{\sum_{i=1}^m r_i}\right)\cdot O(2^n) = O(n^{O(m)})\cdot O(2^n) \\
        &= O(n^{O(1/\alpha)})\cdot O(2^n) = O(n^{O(\alpha n)}).
    \end{align*}
    %It remains to construct finite sets that satisfy the second and third properties. Repeat the quantization trick on the index set $I := \{1,\cdots,\lfloor \alpha n\rfloor\}$, we can show that there exists a set that covers $v_{I}$ within $\delta n^{-\gamma-1}$ is of size at most
    %\begin{align*}
    %    O\left(\left(\frac{1}{qn^{-\gamma-1}\delta}\right)^{\alpha n}\right) = O(n^{O(\alpha n)}).
    %\end{align*}
    For the eigenvalue, we also have there exists a set that covers its domain to be within
    $\delta n^{\gamma}$ with a set of size
    %$\delta n^{-\gamma-1}$ with a set of size
    \begin{align*}
        O\left(\frac{n^{\gamma-1}}{n^{\gamma}\delta}\right) = O(n^{B-1})\le O(n^{O(\alpha n)}).
        %O\left(\frac{n^{\gamma-1}}{q n^{-\gamma-2}\delta}\right) = O(n^{O(\alpha n)}).
    \end{align*}
    with probability at least $1-O(\exp(-\alpha_0 n))$.
    Composing the sets, we find the set $\gN$ has size
    $O(n^{-n/2+O(\alpha n)}q^{-n+O(\alpha n)})$.
    %$O(n^{-n/2+O(\alpha n)}q^{-n})$.
\end{proof}

\begin{lemma}\label{lemma:incompressible_eigenvec_symmetric_markov}
For any eigenvector-eigenvalue pair $(v, \lambda)$ of an \smrm/ $A_n = D_n^{-1}W_n$ with adjacency matrix $W_n$ and the same properties defined in Theorem~\ref{thm:simple_spectrum} and let $(\tilde{v}, \tilde{\lambda}) \in \gN$ be the tuple that well approximates it as defined in Lemma~\ref{lemma:discretize_eigenvec_symmetric_markov}, we have
\begin{align*}
    \|A_{I^c, I}\tilde{v}_I - u\|_2 = O(\delta n^{\gamma}),
    %\|A_{I^c, I}w' - u\|_2 = O(\delta n^{\gamma + 1/2}),
\end{align*}
where $A_{I, J}$ is the matrix formed by row indices from $I$ and column indices from $J$ and $u := (\tilde{\lambda}-A_{I^c, I^c})\tilde{v}_{I^c}$.
\end{lemma}
\begin{proof}
    %Again using the arguments in Lemma~\ref{lemma:compressible_case_symmetric_markov}, we have with probability at least $1-O(\exp(-\alpha_2 n))$,
    %\begin{align*}
        %d_{ii} = n + o(n), 1\leq i\leq n.
    %\end{align*}
    By symmetry, we can let $I=\{1,\cdots,k\}$ for $k=\lfloor\alpha n\rfloor$. Notice by definition we can split $A$ as
    \begin{align*}
    \begin{bmatrix}
    A_k & G \\
    F^\top & A_{n-k}
    \end{bmatrix}
    \begin{bmatrix}
    w\\
    v'
    \end{bmatrix} =
    \lambda
    \begin{bmatrix}
    w\\
    v'
    \end{bmatrix},
    \end{align*}
    where $v=[w^\top,v'^\top]^\top$,
    and as a result,
    \begin{align*}
        &\|F^\top \tilde{v}_I - (\tilde{\lambda} - A_{n-k})\tilde{v}_{I^c}\|_2\\
        \leq & \|F^\top w - (\lambda - A_{n-k})v'\|_2 + \\
        & \|F^\top (\tilde{v}_I-w)\|_2+\|(\tilde{\lambda} - \lambda)\tilde{v}_{I^c}\|_2 +\\
        &\|(\lambda - A_{n-k})(\tilde{v}_{I^c}-v')\|_2 \\
        = & \|F^\top (\tilde{v}_I-w)\|_2 + |(\tilde{\lambda} - \lambda)\tilde{v}_{I^c}\|_2\\
        &+ \|(\lambda - A_{n-k})(\tilde{v}_{I^c}-v')\|_2 \\
        = & O(n^{\gamma-1}\cdot \delta n^{1/2}) + O(n^{\gamma}\delta) \\
        &+ O(n^{\gamma-1}\cdot \delta n^{1/2}) = O(n^{\gamma}\delta).
        %= & O(n^{\gamma-1}\cdot n\cdot n^{-\gamma-1}\delta n^{1/2}) + O(n^{-\gamma-1}\cdot n \cdot\delta n^{1/2}) \\
        %&+ O(n^{\gamma-1}\cdot n\cdot\delta n^{1/2}) = O(n^{\gamma + 1/2}\delta).
    \end{align*}
\end{proof}

Now we are ready to prove Lemma~\ref{lemma:incompressible_case_symmetric_markov}.
\begin{proof}
    Let $\gE$ be the event that there exists some $\delta \in [n^{-B}, n^{-B/2}]$ such that
    \begin{align*}
        n^{-C}\leq \rho_{n^\gamma\delta}(v)\leq n^{0.49}\rho_\delta(v) =: n^{0.49}q
    \end{align*}
    with $q := \rho_{\delta}(v)$ and $\gG$ be the event that 
    \begin{align*}
        \|A_{I^c,I}\tilde{v}_I-u\|_2 = O(\delta n^{\gamma}),
    \end{align*}
    where $u := (\tilde{\lambda} - A_{I^c, I^c})\tilde{v}_{I^c}$ and $(\tilde{v}, \lambda)$ well approximates $(v, \lambda)$ as defined in Lemma~\ref{lemma:incompressible_eigenvec_symmetric_markov}. Let $k := |I| = O(\alpha n)$, from Lemma~\ref{lemma:discretize_eigenvec_symmetric_markov}, we have
    \begin{align*}
        \Pr(\gG^c) = O(\exp(-\alpha_0 n)). 
    \end{align*}
    On the other hand, if $\gE$ occurs,
    define $A_{I^c,I}=[a_{k+1},\ldots,a_n]^\top$, $u=[u_{k+1},\ldots,u_n]^\top$, then
    we have
    \begin{multline*}
        \Pr(\gG|\gE) \leq \sum_{(w', \tilde{v}, \tilde{\lambda})\in\gN}\\
        \Pr\left[\sum_{i=k+1}^{n} |a_i^\top w' - u_i|^2= O(\delta^2 n^{2\gamma + 1})\right]\\
        \leq |\gN|(\rho_{n^\gamma \delta}(v))^{n-k} \leq |\gN|(n^{0.49} q)^{n-k}\\
        =O(n^{-0.01n+O(\alpha n)}),
    \end{multline*}
    which is $O(\exp(-\alpha_0 n))$ if $\alpha$ is chosen small enough.
    As a result, we have
    \begin{align*}
        \Pr(\gE) \leq \Pr(\gG|\gE) + \Pr(\gG^{c}) = O(\exp(-\alpha_0 n)).
    \end{align*}
\end{proof}

\subsection{Finite-sample learnability of \asru/: Unmatched setup}\label{appendix:finite_sample}

\begin{proof}{(Theorem~\ref{thm:discrete_gan_unsup_asr})}
  Under the assumptions that the discriminator is perfect and decomposable and the GAN objective is MMD
  with a linear kernel over the embeddings $D(Y)=\hat{P}^Y$,
  \Eqref{eq:discrete_perfect_discriminator} becomes the following least squares regression problem
  \begin{align}\label{eq:mmd_perfect_discriminator}
    \min_{O'\in \sR^{|\sX|\times|\sY|}} \|\hat{P}^XO'-\hat{P}^Y\|_F^2.
    \end{align}
    Let $\hat{O}$ be the ERM of \Eqref{eq:mmd_perfect_discriminator} and $O$ be the true assignment matrix, by definition and triangle inequality,
    \begin{align*}
      &\|\hat{P}^X\hat{O}-\hat{P}^Y\|_F \\
      &\leq \|\hat{P}^XO-\hat{P}^Y\|_F\\
      &\leq \|\hat{P}^X O-P^Y\|_F+\|\hat{P}^Y -P^Y\|_F.
    \end{align*}
    Apply the triangle inequality again, we have
    \begin{align*}
      &\|\hat{P}^{X}(\hat{O}-O)\|_F \\
      &\leq \|\hat{P}^{X}\hat{O}-\hat{P}^Y\|_F + \|\hat{P}^XO-\hat{P}^Y\|_F\\
      &\leq 2\|\hat{P}^X O-P^Y\|_F+2\|\hat{P}^Y -P^Y\|_F
        %\|\hat{P}^X\hat{O}-\hat{P}^Y\|_F &\leq \|\hat{P}^XO-\hat{P}^Y\|_F\\
        %&\leq \|\hat{P}^X O-P^Y\|_F+\|\hat{P}^Y -P^Y\|_F.
    %\end{align*}
    %Apply the triangle inequality again, we have
    %\begin{align*}
        %\|\hat{P}^{X}(\hat{O}-O)\|_F &\leq \|\hat{P}^{X}\hat{O}-\hat{P}^Y\|_F + \|\hat{P}^XO-\hat{P}^Y\|_F\\
        %&\leq 2\|\hat{P}^X O-P^Y\|_F+2\|\hat{P}^Y -P^Y\|_F
    \end{align*}
    Note that if we replace any $X^{(i)}\rightarrow X^{(i)'}$ and let the resulting empirical distribution be $\hat{P}^{X'}$,
    \begin{align*}
        &\left|\|\hat{P}^X O-P^Y\|_F - \|\hat{P}^{X'}O-P^Y\|_F\right| \\
        \leq & \|(\hat{P}^X-\hat{P}^{X'})O\|_F \leq \frac{\sqrt{2L}}{n^X},
    \end{align*}
    and similarly for $\hat{P}^X$ and $\hat{P}^Y$,
    \begin{align*}
        &\left|\|\hat{P}^X-P^X\|_F - \|\hat{P}^{X'}-P^X\|_F\right| \leq &  \frac{\sqrt{2L}}{n^X}\\
        &\left|\|\hat{P}^Y-P^Y\|_F - \|\hat{P}^{Y'}-P^Y\|_F\right| \leq &  \frac{\sqrt{2L}}{n^Y}.
    \end{align*}
    Therefore, we can apply McDiarmid's inequality to obtain
    \begin{align*}
        &\Pr\left[\|\hat{P}^X - P^X\|_F \geq \frac{\sqrt{L|\sX|}}{\sqrt{n^X}}+\epsilon\right] \leq e^{-\frac{n^X \epsilon^2}{L}}\\
        &\Pr\left[\|\hat{P}^X O - P^Y\|_F \geq \frac{\sqrt{L|\sY|}}{\sqrt{n^X}}+\epsilon\right] \leq e^{-\frac{n^X \epsilon^2}{L}}\\
        &\Pr\left[\|\hat{P}^Y - P^Y\|_F \geq \frac{\sqrt{L|\sY|}}{\sqrt{n^Y}}+\epsilon\right] \leq e^{-\frac{n^Y \epsilon^2}{L}}.
    \end{align*}
    Moreover, let $\epsilon^{XX} := \frac{\sqrt{L|\sX|}}{\sqrt{n^X}}+\epsilon$,
    $\epsilon^{YX} := \frac{\sqrt{L|\sY|}}{\sqrt{n^X}}+\epsilon$,
    $\epsilon^{YY}=\frac{\sqrt{L|\sY|}}{\sqrt{n^Y}}+\epsilon$, then by a union bound, we have
    \begin{multline*}
        \Pr\left[\|\hat{P}^X (\hat{O} - O)\|_F \geq \epsilon^{YX}+\epsilon^{YY}\right] \leq \\ 
        \Pr\left[\|\hat{P}^X \hat{O} - P^Y\|_F + \|\hat{P}^Y - P^Y\|_F \geq \right.\\ \left.\frac{\epsilon^{YX}+\epsilon^{YY}}{2}\right] \\
        \leq  \Pr\left[\|\hat{P}^{YX} \hat{O} - P^{YY}\|_F \geq \frac{\epsilon^{YX}}{2}\right] + \\
        \Pr\left[\|\hat{P}^Y - P^Y\|_F\geq \frac{\epsilon^{YY}}{2}\right] \leq  e^{-\frac{n^X\epsilon^2}{4L}} + e^{-\frac{n^Y\epsilon^2}{4L}}.
    \end{multline*}
    Therefore, we have with probability at least $1-e^{-\frac{n^X\epsilon^2}{4L}} - e^{-\frac{n^Y\epsilon^2}{4L}}$,
    \begin{align*}
        &\epsilon^{YX} + \epsilon^{YY} \geq \|\hat{P}^X(\hat{O}-O)\|_F\\
    %Therefore, we can apply McDiarmid inequality to obtain \textcolor{red}{TODO Verify}
    %\begin{align*}
    %    &\Pr\left[\|\hat{P}^X - P^X\|_F \geq \frac{1}{\sqrt{n^X}}+\epsilon\right] \leq e^{-\frac{n^X \epsilon^2}{L}}\\
    %    &\Pr\left[\|\hat{P}^X O - P^Y\|_F \geq \frac{1}{\sqrt{n^X}}+\epsilon\right] \leq e^{-\frac{n^X \epsilon^2}{L}}\\
    %    &\Pr\left[\|\hat{P}^Y - P^Y\|_F \geq \frac{1}{\sqrt{n^Y}}+\epsilon\right] \leq e^{-\frac{n^Y \epsilon^2}{L}}.
    %\end{align*}
    %Moreover, let $\epsilon^X := \frac{1}{\sqrt{n^X}}+\epsilon$, $\epsilon^Y=\frac{1}{\sqrt{n^Y}}+\epsilon$, then by union bound, we have
    %\begin{multline*}
    %    \Pr\left[\|\hat{P}^X (\hat{O} - O)\|_F \geq \epsilon^X+\epsilon^Y\right] \leq \\ 
    %    \Pr\left[\|\hat{P}^X \hat{O} - P^Y\|_F + \|\hat{P}^Y - P^Y\|_F \geq \frac{\epsilon^X+\epsilon^Y}{2}\right] \\
    %    \leq  \Pr\left[\|\hat{P}^X \hat{O} - P^Y\|_F \geq \frac{\epsilon^X}{2}\right] + \\
    %    \Pr\left[\|\hat{P}^Y - P^Y\|_F\geq \frac{\epsilon^Y}{2}\right] \leq  e^{-\frac{n^X\epsilon^2}{4L}} + e^{-\frac{n^Y\epsilon^2}{4L}}.
    %\end{multline*}
    %Therefore, we have with probability at least $1-e^{-\frac{n^X\epsilon^2}{4L}} - e^{-\frac{n^Y\epsilon^2}{4L}}$,
    %\begin{align*}
    %    &\epsilon^X + \epsilon^Y \geq \|\hat{P}^X(\hat{O}-O)\|_F\\
        &\geq \|P^X(\hat{O}-O)\|_F - \|\hat{P}^X-P^X\|_F\|\hat{O}-O\|_F\\
        &\geq (\sigma_{\min}(P^X)-\|\hat{P}^X-P^X\|_F)\|\hat{O}-O\|_F,
    \end{align*}
    and combined with the bound on $\|\hat{P}^X-P^X\|_F$, we obtain with probability at least $(1-e^{-\frac{n^X\epsilon^2}{4L}} - e^{-\frac{n^Y\epsilon^2}{4L}})(1-e^{-\frac{n^X\epsilon^2}{4L}})$,
    \begin{align*}
        \|\hat{O}-O\|_F \leq \frac{\epsilon^{YX}+\epsilon^{YY}}{\sigma_{\min}(P^X)-\epsilon^{XX}}.
    \end{align*}
    Assume the correct mapping is deterministic, so that $O_{xy}\in\{0,1\}$ and each row
    has only one nonzero element, then
    to achieve perfect \asru/, we need for any $x\in \sX$ and $y\neq G(x)$,
    %    \|\hat{O}-O\|_F \leq \frac{\epsilon^X+\epsilon^Y}{\sigma_{\min}(P^X)-\epsilon^X}.
    %\end{align*}
    %To achieve perfect \asru/, we need for any $x\in \sX$ and $y\neq G(x)$,
    \begin{align*}
         &|\hat{O}_{xG(x)}-\hat{O}_{xy}|  > 0\\
        &\Longleftarrow  1-|\hat{O}_{xG(x)}-O_{xG(x)}|-|\hat{O}_{xy}-O_{xy}| > 0\\
        &\Longleftarrow 1-2\|\hat{O}-O\|_{\infty} > 0 \Longleftrightarrow \|\hat{O}-O\|_F < \frac{1}{2},
    \end{align*}
    which occurs if
    \begin{align*}
        \sigma_{\min}(P^X) > \epsilon^{XX}+2\epsilon^{YX}+2\epsilon^{YY}.
    \end{align*}
    \end{proof}
    
    %    \sigma_{\min}(P^X) > 3\epsilon^X+2\epsilon^Y.
    %\end{align*}
    %\end{proof}

\subsection{Training dynamic of \asru/}
To prove Theorem~\ref{thm:train_dynamic}, we need the following lemma on the properties of the gradient of the softmax function based on 
\cite{Gao2017-softmax}.
\begin{lemma}\label{lemma:softmax}
Let $H(x)$ be the Jacobian matrix of the softmax function $\sigma: \sR^d \mapsto \sR^d$ with $\sigma_i(x)=\frac{e^{x_i}}{\sum_{j=1}^d e^{x_j}}$, then we have $H(x) = \diag(\sigma(x))-\sigma(x)\sigma(x)^\top$ and $H(x)$ is positive semi-definite (PSD) with the null space $\mathrm{span}\{\mathbf{1}_d\}$.
\end{lemma}
\begin{proof}
    Apply product rule of calculus, we have
    \begin{align*}
        H_{ij}(x)&=\frac{\partial \sigma_i(x)}{\partial x_j} \\
        &= \delta_{ij}\sigma_i(x) - \frac{e^{x_i}e^{x_j}}{(\sum_{j=1}^d e^{x_j})^2}\\
        &= \delta_{ij}\sigma_i(x) -\sigma_i(x)\sigma_j(x),
    \end{align*}
    and therefore $H(x)=\diag(\sigma(x)) - \sigma(x)\sigma(x)^\top$. To show that $H(x)$ is PSD, notice that
    \begin{align*}
        v^\top H(x)v &= v^\top \diag(\sigma(x))v - (v^\top \sigma(x))^2\\
        &= \E_{I\sim\sigma(x)}[v_{I}^2] - \E_{I\sim\sigma(x)}^2[v_I]\\
        &= \Var{(v_I)} \geq 0,
    \end{align*}
    where by Jensen's inequality, achieves ``$=$'' if and only if $v_i=\sigma^\top v = C,\,\forall i$ for some constant $C$.
\end{proof}
Next, we shall establish explicit formula for NTKs of the discriminator and the generator. For clarity, we will copy the formula for the discriminator and the generator used in our analysis:
\begin{align}
    f_{\tau, l}(y) &= \lim_{m\rightarrow \infty}\frac{1}{\sqrt{m}}\sum_{r=1}^m v^{\tau, l}_r\max\{W^{\tau, l}_{ry}, 0\},\label{eq:discrim_dnn2}\\
    P^{g_{t}}_l(y) &= \E_{X\sim P^X_l}\left[O_t(y|X)\right]\nonumber\\
    &:= \E_{X\sim P^X_l}\left[\frac{\exp(U_{y}^{t\top} x)}{\sum_{y'\in \sY}\exp(U_{y'}^{t\top} x)}\right] \label{eq:gen_dnn2}.
\end{align}
\begin{lemma}\label{lemma:discrim_ntk}
For the NTKs of the discriminators defined by \Eqref{eq:discrim_dnn2}, we have $K_{D,l} \equiv K_{D,1},\,1\leq l \leq L$ and $\mathbf{1}_{|\sY|}$ is an eigenvector of $K_{D,1}$.
\end{lemma}
\begin{proof}
    For simplicity, we ignore the dependency on $\tau$ for the terms in the proof. First, by definition, we have
    \begin{align*}
        \frac{\partial f_l(y)}{\partial W_r^l} &= \lim_{m\rightarrow \infty}\frac{1}{\sqrt{m}}\sum_{r=1}^m v_r^l e_y \mathbbm{1}[W_{ry}^l \geq 0],\\
        \frac{\partial f_l(y)}{\partial v_r^l} &= \lim_{m\rightarrow \infty} = \frac{1}{\sqrt{m}}\max\{W_{ry}^l, 0\}
    \end{align*}
    and therefore
    \begin{align*}
        & \E_{v^l,W^l\sim \gN(0, I)}\left[\frac{\partial f_l(y)}{\partial W_r^l}^\top\frac{\partial f_l(y)}{\partial W_r^l}\right] = \\
        &\lim_{m\rightarrow \infty}\frac{1}{m}\E_{v^l, W^l\sim \gN(0, I)}\sum_{r=1}^m \delta_{yy'} v_r^2\mathbbm{1}[W_{ry}^l \geq 0]\\
        &= \delta_{yy'}\frac{1}{m}\sum_{r=1}^m\E_{W_{ry}^l\sim \gN(0, 1)}[\mathbbm{1}[W_{ry}^l \geq 0]]\\
        &= \frac{1}{2}\delta_{yy'}.
    \end{align*}
    On the other hand,
    \begin{align*}
        & \E_{v^l,W^l\sim \gN(0, I)}\left[\frac{\partial f_l(y)}{\partial v^l}^\top\frac{\partial f_l(y')}{\partial v^l}\right]\\
        =& \frac{1}{m}\E_{v^l,W^l}\left[\sum_{r=1}^m\max\{W^l_{ry}, 0\}\max\{W^l_{ry'}, 0\}\right]\\
        =& \begin{cases}
        \E_{v^1_1,W^1_1}\left[\max\{W_{11}^1, 0\}^2\right]\quad\text{ if }y=y',\\
        \E_{v^1_1,W^1_1}\left[\max\{W_{11}^1, 0\}\right]^2\quad\text{ otherwise}.
        \end{cases}
    \end{align*}
    Therefore,
    \begin{align*}
        &K_{D, l}(y, y') = \\
        &\begin{cases}
        \left(\frac{1}{2}+ \E_{v^1_1,W^1_1}\left[\max\{W_{11}^1, 0\}^2\right]\right)\quad\text{if y=y',}\\
        \E_{v^1_1,W^1_1}\left[\max\{W_{11}^1, 0\}\right]^2\quad\text{ otherwise}.
        \end{cases}
    \end{align*}
    Notice that the sum of every row in $K_{D, l}$ is
    \begin{multline*}
        \left(\frac{1}{2}+ \E_{v^1_1,W^1_1}\left[\max\{W_{11}^1, 0\}^2\right]\right) +\\ (|\sY|-1)\E_{v^1_1,W^1_1}\left[\max\{W_{11}^1, 0\}\right]^2,
    \end{multline*}
    and thus $\mathbf{1}_{|\sY|}$ is an eigenvector of $K_{D, l}$.
\end{proof}

\begin{lemma}\label{lemma:gen_ntk}
For the generator defined by \Eqref{eq:gen_dnn2}, we have
\begin{align}\label{eq:ntk_gen_explicit}
    &K_{O_{t, x}} = \nonumber\\
    &\E_{U_{1:|\sY|}\sim \gN(0, I)}\left[(\diag(O_x)-O_xO_{x}^{\top})^2\right].
\end{align}
Further, the null space of $K_{O_{t, x}}$ is $\mathrm{span}\{\mathbf{1}_{|\sY|}\}$.
\end{lemma}
\begin{proof}
    For simplicity, we ignore the dependency on $t$ for the terms in the proof. By chain rule,
    \begin{align*}
        &\frac{\partial O_x(y)}{\partial U_{x'
        y'}} = \frac{\partial h_{y'}(x)}{\partial U_{xy'}}\frac{\partial O_x(y)}{\partial h_{y'}(x)}\\
        &= \delta_{xx'}(O(y|x)\delta_{yy'}-O(y|x)O(y'|x))
    \end{align*}
    As a result,
    \begin{align*}
    &\sum_{d, y'}\frac{\partial O_x(y)}{\partial U_{dy'}}^\top\frac{\partial O_x(y'')}{\partial U_{dy'}}\\
    &= \sum_{y'}(O_x(y)\delta_{yy'}-O_x(y)O_x(y'))\\
    &(O_x(y'')\delta_{y''y'}-O_x(y'')O_x(y'))\cdot\\
    &= ((\diag(O_x)-O_xO_x^\top)^2)_{yy''}
    \end{align*}
    Take the expectation over $U$ and put everything in matrix form, we obtain
    \begin{align*}
        K_{O_x} = \E_{U\sim \gN(0, I)}\left[(\diag(O_x)-O_xO_x^\top)^2\right].
    \end{align*}
    Next we shall study the null space of $K_{O_x}$. From Lemma~\ref{lemma:softmax}, we have $H_x:=\diag(O_x)-O_xO_x^\top$ is PSD with null space $\mathrm{span}\{\mathbf{1}_{|\sY|}\}$, and thus
    \begin{align*}
        &v^\top K_{O_x}v = \E_{U\sim \gN(0, I)}\left[\|H_x v\|^2\right] \geq 0,
    \end{align*}
    with equality achieved if and only if
    \begin{align*}
        H_x v = 0,\forall x\in\sX \Leftrightarrow v \in \mathrm{span}(\mathbf{1}_{|\sY|}). 
    \end{align*}
\end{proof}
We are now ready to prove Theorem~\ref{thm:train_dynamic}.
\begin{proof}{(Theorem~\ref{thm:train_dynamic})}
    When the objective is MMD, the discriminator can be decomposed as 
    $$
    a_{f_\tau}(y)=f_\tau(y)=\sum_{l=1}^L f_{\tau, l}(y_l),
    $$
    we have
    \begin{align}
        \Ls_t(f) = \sum_{l=1}^L \E_{Y_l\sim P^Y_l}[f_l(Y_l)] - \E_{Y_l'\sim P^X_l O_t}[f_l(Y_l')],
    \end{align}
    and the discriminator dynamic PDE \Eqref{eq:ntk_pde_decomp_discrim_gd} becomes:
    \begin{align*}
        &\partial_\tau f_{\tau, l} = K_{D, l} (P^Y_l-P^X_l O_t)^\top.
    \end{align*}
    Without much loss of generality, suppose we initialize $f_{0, l}(y)\equiv 0$ and stop training the discriminator after $\tau_{\max}$ steps. The solution for the  discriminator PDE is then simply
    \begin{align}\label{eq:optimal_mmd_discrim}
        f_{g_t, l} = \tau_{\max} K_{D, l} (P^Y_l-P^X O_t)^\top.
    \end{align}
   Plug this expression into the generator loss and apply Lemma~\ref{lemma:discrim_ntk}, we obtain
   \begin{align*}
        \Cs_t(g_t) &:= \tau_{\max}\sum_{l=1}^L\|P^Y_l - P^X_l O_t\|_{K_{D, l}}^2\\
        &= \tau_{\max}\|P^Y - P^X O_t\|_{K_{D, 1}}^2,
    \end{align*}
    where $\|A\|_{K} = \sqrt{\Tr(A K A^\top)}$ is the kernelized norm of $A$ by kernel $K$.
    
    Further, plug \Eqref{eq:optimal_mmd_discrim} into the generator PDE \Eqref{eq:ntk_pde_decomp_gen_gd}, we obtain
    \begin{align*}
        \partial_t O_{t,x}^\top &= K_{O_{t,x}}\sum_{l=1}^L P^X_l(x)K_{D, l}(P_l^Y-P_l^XO_t)^\top\\
        &= K_{O_{t,x}}K_{D,1}(P^Y - P^XO)^\top \tilde{P}^X_x,
    \end{align*}
    where $\tilde{P}^X_x$ is the $x$-th column of $P^X$. Next, notice that
    \begin{align*}
        &\frac{\partial \Cs_t}{\partial O_{t, xy}} \\
        =& 2\tau_{\max} K_{D,1}(y, \cdot) (P^X O - P^Y)^\top \tilde{P}_x^X\\
        \Longrightarrow & \frac{\partial \Cs_t}{\partial O_t} = P^{X\top} (P^X O - P^Y)K_{D,1}.
    \end{align*}
    Then apply the chain rule,
    \begin{multline*}
        \partial_{t}\Cs_t = \Tr\left(\frac{\partial \Cs_t}{\partial O_t}^\top \frac{\partial O_t}{\partial t}\right) \\
        = \sum_{x\in\sX}\Tr\left(\frac{\partial \Cs_t}{\partial O_{t,x}} \frac{\partial O_{t,x}}{\partial t}^\top\right) = \\
        -\tau_{\max}\sum_{x\in\sX}\|\tilde{P}^{X\top}_x (P^Y-P^X O_t)\|_{K_{D, l}K_{G, l} K_{D, l}}^2.
    \end{multline*}
    Now, apply Lemma~\ref{lemma:discrim_ntk}, we 
    have
    \begin{align*}
         & \partial_\tau f_{\tau, l}^\top \mathbf{1}_{|\sY|}\\
        =& (P^Y_l-P^X_lO_t) K_{D,l} \mathbf{1}_{|\sY|}\\
        =& \lambda(P^Y_l-P^X_lO_t)\mathbf{1}_{|\sY|} = 1-1 = 0\\
        \Longrightarrow & \mathbf{1}_{|\sY|} \perp K_{D,l}(P^Y_l-P^X_l O_t)^\top,
    \end{align*}
    where $\lambda$ is the eigenvalue of $K_{D,l}$ associated with $\mathbf{1}_{|\sY|}$, and thus \begin{align*}
         K_{D,l}(P^Y-P^X O_t)^\top\tilde{P}_x^X \perp \mathbf{1}_{|\sY|}.
    \end{align*}
    As a result, using Lemma~\ref{lemma:gen_ntk}, we conclude that the kernelized residual vector $\partial_\tau f_{\tau, l}$ is always perpendicular to the null space of the stepwise generator NTK $K_{O_{t, x}}$ for all $1\leq l\leq L,\,x\in\sX$, and thus 
    \begin{align*}
        &\|K_{D,l}(P^Y-P^X O_t)^\top\tilde{P}_x^X\|_{K_{G,l}} \\
        \geq& \lambda_G\|K_{D,l}(P^Y-P^X O_t)^\top\tilde{P}_x^X\|_2\\
        \geq& \lambda_G\lambda_D \|P^Y-P^X O_t\|_{K_{D,1}},
    \end{align*}
    where 
    \begin{align*}
        \lambda_{G} &\geq \min_{1\leq l\leq L}\lambda_{|\sY|-2}(K_{G, l}) > 0,\\
        \lambda_{D} &\geq \lambda_{\min}(K_{D, 1}) > 0.
    \end{align*}
    Summing over $x$, we obtain
    \begin{align*}
        \partial_t \Cs_t \leq -\tau_{\max} \lambda_G\lambda_D \|P^{X\top} (P^Y-P^XO_t)\|^2_{K_{D,1}}.
    \end{align*}
    Under the assumption that $P^XO=P^Y$ has at least one solution, we have $P^Y-P^X O$ is in the range space of $P^X$, which implies 
    \begin{multline*}
        \|P^{X\top} (P^Y-P^XO_t)\|^2_{K_{D,1}} \geq \\
        \lambda_{X} \|P^Y-P^XO_t\|^2_{K_{D,1}},
    \end{multline*}
    for some $\lambda_X > 0$. Put together the results, we can bound the convergence rate of the generator loss by  
    \begin{align*}
        &\partial_t \Cs_t \leq -\tau_{\max}\lambda_{G}\lambda_D\lambda_X \Cs_t\\
        \Longrightarrow & \Cs_t \leq \Cs_0 e^{-\tau_{\max}\lambda_{G}\lambda_D\lambda_X t}\xrightarrow{t\rightarrow\infty}0,
    \end{align*}
    which implies that $\lim_{t\rightarrow\infty}P^X O_t = P^Y$.
    
    % Now, assuming the generator is linear and $O_t(x) := W^\top x$ and let $O_t(x) := (W_{xy})_{y\in \sY}$, then we have
    % \begin{align*}
    %    K_{G, l}&=\sum_{x\in \sX} P^X_l(x)(\diag(O_t(x))-O_t(x)O_t(x)^\top)\\
    %    &=: \sum_{x\in \sX}P^X_l(x)H_t(x).
    % \end{align*}
    % Then by Lemma~\ref{lemma:softmax}, we have $H_t$ is PSD with a null space in the direction $\mathbf{1}_d$. As a result, we can show that $K_{G, l}$ also has a null space in the direction $\mathbf{1}_d$. This can be seen by the fact that $P_l^X(x) > 0, \forall x \in \sX$ and
    %\begin{align*}
    %    v^\top K_{G,l}v &= \sum_{x\in \sX}P_l^X(x)v^\top H_t(x)v\geq 0,
    %\end{align*}
    %where equality is achieved if and only if $v^\top H_t(x)v = 0, \forall x\in \sX$, which implies $v =\mathbf{1}_d$.  
\end{proof}

%% file: appendix_reproducibility.tex
\section{Reproducibility checklist}\label{appendix:reproducibility}
\paragraph{Synthetic language creation} To create a synthetic HMM language, we need to specify the initial probability vector $\pi$, the transition probability matrix $T$, the generator matrix $O$ and the maximal length of the utterances $L$. 

\emph{Initial probability}: we create $\pi$ by first uniformly randomly sampling each coefficient between $[0, 1]$ and then normalizing the resulting vector by its sum.

\emph{Transition probability}:
for the asymptotic setting, for all three languages, we control the number of eigenvalues $m$ of its transition matrix using a disjoint union of identical sub-graphs with $m$ eigenvalues, with the remainder of the nodes being self-loops. The parameters and the procedure used to determine them are as follows:
\begin{itemize}
    \item \emph{Circulant graph}: only undirected cycles or equivalently, circulant graph with the action set $\{-1, 1\}$, are used. Since the distinct eigenvalues of an undirected $n$-cycle $C_n$ are $-\cos\left(\frac{2\pi k}{n}\right),k=0,\cdots,\lfloor \frac{n-1}{2}\rfloor+1$, we can create a Markov graph with $|\sX|^N$ nodes and $n\pm 1$ eigenvalues by a disjoint union of $\lfloor\frac{|\sX|^N}{2n-1}\rfloor$ $C_{2n-1}$ graphs. In our phase transition experiment, we fix $N=2$ and vary $10\leq|\sX|\leq 14$ and $2\leq n\leq 20$;
    \item \emph{De Bruijn graph}: an undirected \debruijn/ graph $\mathrm{DB}(k, m)$ is a graph with $k^m$ nodes such that node $i$ connects to any node $j$ whose $k$-ary numerals $v(i)$ and $v(j)$ satisfies $v_{2:m}(i)=v_{1:m-1}(j)$. Clearly, $m$ is the in/out-degree of the graph. The eigenvalues of $\mathrm{DB}(k, m)$ are known to be $\cos\left(\frac{i\pi}{j}\right), 0\leq i < j\leq m+1$ \cite{Delorme1998-debruijn-spectrum}. Therefore, we can create a Markov graph with $|\sX|^N$ nodes and at most $n$, $n\leq (\lfloor \log_k |\sX|^N\rfloor+1)^2/2$ distinct eigenvalues by a disjoint union of $\frac{|\sX|^N}{k^{\sqrt{2m}-1}}$  $\mathrm{DB}(k, \sqrt{2n}-1)$ graphs. For the phase transition experiment, we set the in/out-degree of the \debruijn/ subgraphs to be $2$ and the $N$-gram size $N=3$, and we vary $8\leq |\sX|\leq 11$ and $2\leq n \leq 32$ with a step size of $2$ for the latter.
    \item \emph{Hypercube}: an $n$-cube $Q_n$ is a graph with $2^n$ nodes such that node $i$ connects to any node $j$ with Hamming distance between their binary numerals $d_H(b(i), b(j))=1$. The eigenvalues of the adjacency matrix of $Q_n$ is $1-\frac{2k}{n},k=0, \cdots, n$. Therefore, we can create a Markov graph with $|\sX|^N$ nodes and $n\leq \lfloor N \log_2 |\sX| \rfloor$ eigenvalues by a disjoint union of $\lfloor\frac{|\sX|^N}{2^n} \rfloor$ $n$-cubes. For the phase transition experiment, we fix $N=4$, and vary $5\leq |\sX|\leq 8$ and $2 \leq n\leq 9$.
\end{itemize}

In the finite-sample setting, we create transition matrices for phase transition experiments using two different setups:
\begin{itemize}
    \item For the circulant graph, we vary its action set to be $\{1,\cdots,d\}$, where $d$ takes values from $2$ to $81$ with a step size of 8;
    \item For the other two graphs, we linearly interpolate between the underlying graph $T_G$ and its Hamiltonian cycle $T_C$ as
    \begin{align}\label{eq:finite_sample_graph}
        T = (1-w) T_G + w T_C,
    \end{align}
    with a weight $w \in [0, 1]$. In particular, for the \debruijn/ graph, the weight for the cycle $w$ takes 10 different values equally spaced between $[0, 1]$; for the $n$-cube, the weight $w$ takes 10 different values equally spaced between $[0.98, 1]$.
\end{itemize}
 
 \emph{Generator matrix} $O$: set by assuming $|\sX|=|\sY|$ and randomly permuting the rows of the $|\sX|\times|\sX|$ identity matrix.

 \emph{Sampling}: in the asymptotic case, no sampling is needed and we simply set maximal length $L=20$ for cycle graph and 10 for the other two graphs. For the finite-sample case, the synthetic speech and text datasets are created independently by sampling from the same HMM twice. For all three graphs, we sample $n^X=n^Y=2560$ utterances for both text and speech with $L=40$ for the \debruijn/ graph and $L=80$ for the other two graphs.

\paragraph{Model architecture} We use a one-layer linear generator with $|\sX|$ input nodes and $|\sY|$ output nodes, with no bias. Next, for all experiments except the experiment on different generator averaging strategies,  we use a one-layer CNN with $|\sY|$ input channels, 1 output channel and a $1\times L$ kernel with no bias. For the experiment on different averaging strategies, we use instead a sequence of 2-layer MLPs with 128 hidden nodes and ReLU activation function, one at each time step, as the discriminators. For all experiments, we disable the logits for special tokens and silences during training and testing. 

\paragraph{Training setting} SGD with a learning rate of 1.0 is used to train the discriminator, while Adam with a learning rate of 0.005 is used to train the generator. The dataset is used as a single batch for all experiments, though we do not observe any significant drop in performance using smaller batch sizes. No weight decays or dropouts is used. Further, we alternatively train the generator and discriminator 1 epoch each, and reset the discriminator weight to 0 for the linear case and to random Gaussian weights using Xavier initialization in the nonlinear case. All experiments are conducted on a single 12GB NVIDIA GeForce GTX 1080Ti GPU.